\documentclass[reprint,aps,pre,superscriptaddress]{revtex4-2}
\usepackage{amsmath}
\usepackage{amsthm}
\usepackage{fancyhdr}
\usepackage{amsfonts}
\usepackage{amssymb}
\usepackage{bm}
\usepackage{enumitem}
\usepackage{upgreek}
\usepackage{graphicx}
\usepackage{caption}
\usepackage{dsfont}
\graphicspath{{figures/}}
\usepackage{epstopdf}
\usepackage{hyperref}
\usepackage{xcolor}

\newcommand{\pderiv}[2]{\frac{\partial {#1}}{\partial {#2}}}

\let\originalleft\left
\let\originalright\right
\renewcommand{\left}{\mathopen{}\mathclose\bgroup\originalleft}
\renewcommand{\right}{\aftergroup\egroup\originalright}

\usepackage{xcite}
\usepackage{xr-hyper}
\usepackage{hyperref}

\makeatletter
\newcommand*{\addFileDependency}[1]{
  \typeout{(#1)}
  \@addtofilelist{#1}
  \IfFileExists{#1}{}{\typeout{No file #1.}}
}
\makeatother

\newcommand*{\myexternaldocument}[1]{%
    \externaldocument{#1}%
    \addFileDependency{#1.tex}%
    \addFileDependency{#1.aux}%
}

\myexternaldocument{CMIPS_SI_v1.9}

\begin{document}
\title{Chemotactic motility-induced phase separation}

\author{Hongbo Zhao}
\affiliation{Department of Chemical and Biological Engineering,\\Princeton University, Princeton, NJ 08544}
\author{Andrej Ko\v{s}mrlj}
\affiliation{Department of Mechanical and Aerospace Engineering, \\Princeton University, Princeton, NJ 08544}
\affiliation{Princeton Materials Institute, \\Princeton University, Princeton, NJ 08544}
\author{Sujit S. Datta}
\email{ssdatta@princeton.edu}
\affiliation{Department of Chemical and Biological Engineering,\\Princeton University, Princeton, NJ 08544}
\date{\today}

\begin{abstract}
Collectives of actively-moving particles can spontaneously separate into dilute and dense phases—a fascinating phenomenon known as motility-induced phase separation (MIPS).  MIPS is well-studied for randomly-moving particles with no directional bias. However, many forms of active matter exhibit collective chemotaxis, directed motion along a chemical gradient that the constituent particles can generate themselves. Here, using theory and simulations, we demonstrate that collective chemotaxis strongly competes with MIPS---in some cases, arresting or completely suppressing phase separation, or in other cases, generating fundamentally new dynamic instabilities. We establish quantitative principles describing this competition, thereby helping to reveal and clarify the rich  physics underlying active matter systems that perform chemotaxis, ranging from cells to robots. 
\end{abstract}

\maketitle

The thermodynamics of active matter---collections of active agents that consume energy---has been studied extensively due to its fundamental richness as well as its importance to biological and engineering applications~\cite{Marchetti2013,Gompper2020}. One prominent class of active matter is that composed of self-propelled agents, ranging from enzymes~\cite{Mohajerani2018,Agudo-Canalejo2018,Jee2018}, motile microorganisms~\cite{Murray2007,Liu2019}, and mammalian cells~\cite{Alert2019,Scarpa2016} to  synthetic microswimmers and robots~\cite{Palacci2013,Theurkauff2012,Palagi2018}. These forms of active matter can often be modeled as collections of Active Brownian Particles (ABPs), each of which self-propels with a velocity of magnitude $U_0$ and a direction that is continually reoriented by random thermal fluctuations, eventually decorrelating over a time scale $\tau_R$. The persistence length of an ABP trajectory is then given by $\sim U_0\tau_R$; for a particle of radius $a$, its directedness can therefore be described by the reorientation P\'{e}clet number $\text{Pe}_\text{R} \equiv a / (U_0\tau_R)$.

Studies of this canonical model have led to fascinating insights into the nonequilibrium thermodynamics of active matter. For example, phase separation in passive equilibrium systems typically requires attractive interactions between the constituents; in stark contrast, for sufficiently small $\text{Pe}_\text{R}$, collections of ABPs undergo motility-induced phase separation (MIPS) into dense and dilute phases without requiring attractive interactions~\cite{Redner2013a,Fily2012,Cates2015,Takatori2015}.  Even more surprisingly, despite this process being highly out-of-equilibrium, its spatiotemporal dynamics can in some cases be described using models inspired by the classical Cahn-Hilliard theory of phase separation of thermally-equilibrated passive systems ~\cite{Stenhammar2013,Tjhung2018,Speck2014,Cates2015,Cates2010}.

This prior work focused on ABPs that move randomly, with no preferred direction. However, many examples of active matter exhibit collective \emph{chemotaxis}---directed motion in response to an external chemical gradient that can be generated collectively by the agents themselves. In biology, this phenomenon enables populations of cells to escape from harmful environments, colonize new terrain, and migrate as groups~\cite{Murray2007,Berg1975,Cremer2019,Fu2018,Bhattacharjee2019,Bhattacharjee2019a,Bhattacharjee2021,Bhattacharjee2022}; at the subcellular level, enzymes may also perform chemotaxis~\cite{Mohajerani2018,Agudo-Canalejo2018,Jee2018}. Synthetic forms of active matter that can perform chemotaxis have also been developed. Studies using these model systems have revealed new surprises in their phase behavior---e.g., unusual clustering and oscillatory density fluctuations that are not captured by current models of
MIPS~\cite{Palacci2013,Theurkauff2012,Stark2018,Pohl2014,Liebchen2018,Liebchen2015,Liebchen2017,Liebchen2017a,Saha2014,Saha2019,Varga2022,Varga2022pre}. However, despite these hints that chemotaxis can influence the physics of active matter, a broader understanding of how exactly chemotaxis alters MIPS remains lacking.

Here, we address this gap in knowledge by developing a theoretical model that combines both MIPS and chemotaxis, which are usually studied in isolation. We find that collective chemotaxis can dramatically suppress MIPS, arrest phase separation, or engender new complex phase separation dynamics---as controlled by the competition between MIPS, which drives ABPs to cluster into dense phases, and chemotaxis, which instead drives them to disperse away. Our analysis of this competition establishes quantitative principles describing how chemotaxis influences MIPS, thereby expanding current understanding of its rich phenomenology.

\textbf{\emph{Governing equations.}}
Building on existing continuum models of MIPS~\cite{Stenhammar2013,Tjhung2018,Speck2014,Cates2015,Cates2010}, we describe the time evolution of the volume fraction $\phi$ of \emph{chemotactic} ABPs via the continuity equation,
\begin{gather}
  \pderiv{\phi}{t} = -\nabla \cdot \mathbf{J}, \label{eqn::dphidt} \\
  \mathbf{J} = \underbrace{- M_0 \phi \nabla \left( \tilde{\mu}_h(\phi, \text{Pe}_\text{R}) - \kappa \nabla^2 \phi \right)}_{\text{MIPS}} + \underbrace{\chi_0 \phi \nabla f(\tilde{c})}_{\text{chemotaxis}},  \label{eqn::phi_J}
\end{gather}
where $t$ is time and $\mathbf{J}$ is the flux of particles. This flux has two contributions, as indicated by the underbraces in Eq.~\eqref{eqn::phi_J}.
The first reflects active Brownian motion, as established by the classical Cahn-Hilliard model of MIPS (referred to as ``model B'' in the literature); in future work, it would be interesting to explore other models of MIPS that treat additional complexities~\cite{Tjhung2018}. As detailed in Sec.~\ref{sec::SI_ABP} in~[SI], $M_0=0.5U_0^2 \tau_R$ is the active diffusivity reflecting the random undirected motion of the particles, $\tilde{\mu}_h$ is the bulk chemical potential nondimensionalized by the energy scale $0.5\zeta U_0^2 \tau_R$, where $\zeta$ is the drag coefficient, and the characteristic length scale $\sqrt{\kappa} \sim U_0 \tau_R$ sets the width of the interface between the dense and dilute phases in MIPS~\cite{Stenhammar2013,Cates2015}.

The second term in Eq.~\eqref{eqn::phi_J} represents a new addition of chemotaxis to this classical model of MIPS. Here, $\tilde{c}$ is the concentration, nondimensionalized by a fixed characteristic concentration, of a diffusible chemical signal (the \emph{chemoattractant}) that the particles sense and direct their motion in response to. The function $f(\tilde{c})$ describes the ability of the particles to sense the chemoattractant, and typically increases monotonically with $\tilde{c}$; as an illustrative example, we take $f(\tilde{c})=\tilde{c}$ as is often done for simplicity~\cite{Brenner1998,Herrero1996}. The chemotactic coefficient $\chi_0$ describes the ability of the particles to move up the sensed chemoattractant gradient. Thus, $\chi_0 \nabla f(\tilde{c})$  describes the chemotactic velocity, and when multiplied by $\phi$ describes the chemotactic flux~\cite{Keller1971,Keller1971a}. Hence, we define a new chemotactic P\'{e}clet number $\text{Pe}_\text{C} \equiv \chi_0 / M_0$ to describe the competition between directed chemotaxis and undirected active diffusion.

Chemoattractants (e.g., nutrients) are often taken up by the particles themselves---thereby collectively generating a local chemoattractant gradient that the particles bias their motion in response to~\cite{Murray2003,Colin2021,Adler1966,Fu2018,Cremer2019,Bhattacharjee2021,Bai2021,Liebchen2017,Pohl2014,Saha2014,Saha2019}. Thus, we describe the chemoattractant via
\begin{equation} \label{eqn::dcdt}
  \pderiv{\tilde{c}}{t} = D_c \nabla^2 \tilde{c} - k \phi g(\tilde{c}) + S,
  \end{equation}
where $D_c$ is the chemoattractant diffusivity, $k$ is the characteristic rate of chemoattractant uptake per particle, and $g(\tilde{c})$ describes how uptake rate increases with $\tilde{c}$; while $g(\tilde{c})$ is often described by Michaelis-Menten kinetics, here we use the linearized form $g(\tilde{c})=\tilde{c}$ for simplicity. Finally, $S$ represents the rate at which chemoattractant is externally supplied, which we take to be constant and spatially uniform as an illustrative example.

\begin{figure*}[hbt]
  \includegraphics[width=\textwidth]{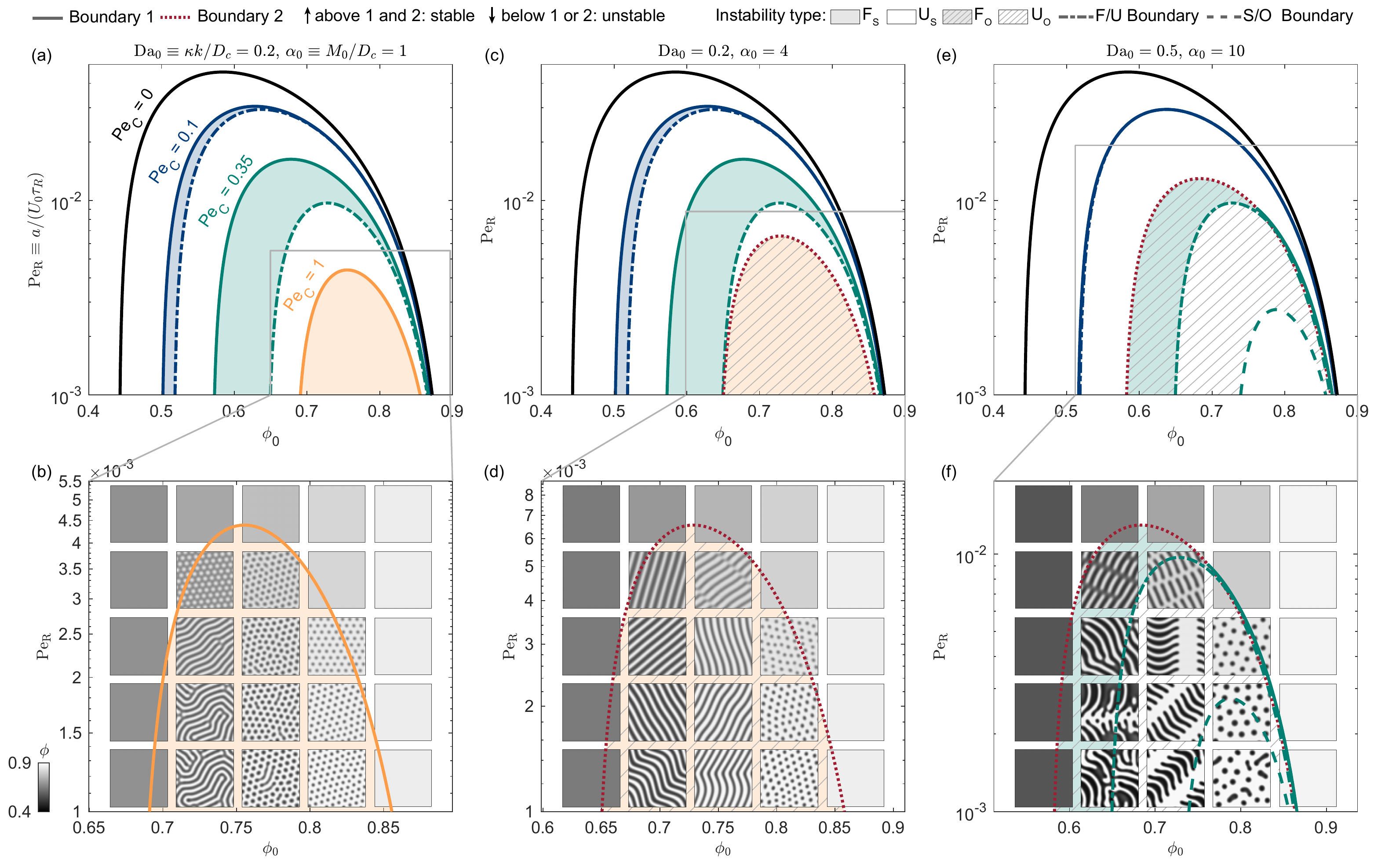}
  \caption{Chemotaxis suppresses MIPS. (a, c, e) Phase diagram, which is typically parameterized by $\phi_0$ and $\text{Pe}_\text{R}$, as predicted by linear stability analysis for different $\text{Da}_0$ and $\alpha_0$. The black curve shows the limit of stability without chemotaxis, below which we observe conventional MIPS. The colored solid and red dotted curves show Boundaries 1 and 2, which are defined in the main text; different colors indicate different values of $\text{Pe}_\text{C}$. Boundary 2 is below the horizontal axis in (a). The region above \emph{both} Boundaries is stable, with ABPs in the homogeneous state, while the region below \emph{either} Boundary is unstable. The different instability types---finite (F) or unbounded (U), stationary (S) or oscillatory (O)---are denoted by the shaded, unshaded, non-hashed, and hashed regions, respectively. Dash-dotted and dashed curves indicate the boundaries between F/U and S/O instabilities, respectively. The linear stability analysis predictions are corroborated by full numerical simulations (Movies S2-S4), snapshots of which are shown in (b, d, f), which focus on the grey boxed regions shown in (a, c, e).}
  \label{fig::ABP}
\end{figure*}

\textbf{\emph{Chemotaxis suppresses MIPS.}}
First, we establish the conventional case of MIPS as a baseline, described by our governing Eqs.~\eqref{eqn::dphidt}--\eqref{eqn::dcdt} in the absence of chemotaxis ($\text{Pe}_\text{C}=0$). To do so, we choose a functional form for $\tilde{\mu}_h(\phi, \text{Pe}_\text{R})$, given by Eq.~\eqref{eqn::mu_h} in~[SI], that derives from a previously-established ABP equation of state~\cite{Takatori2014,Takatori2015}. The homogeneous state with constant, spatially-uniform $\phi(\textbf{x})=\phi_0$, where $\textbf{x}$ denotes position, becomes unstable to fluctuations in $\phi$ when the free energy is nonconvex ($\partial_\phi\tilde{\mu}_h<0$). Therefore, the spinodal curve that demarcates the limit of stability is given by $\partial_\phi\tilde{\mu}_h=0$, shown by the black curves in the $\text{Pe}_\text{R}-\phi_0$ phase diagrams in Fig.~\ref{fig::ABP}, where $\phi_0$ represents the ABP volume fraction averaged over the entire system. Above this spinodal curve, the homogeneous state is linearly stable. Below the spinodal, ABPs spontaneously separates into dense and dilute phases via spinodal decomposition (Movie~S1)---initially forming domains with a characteristic most unstable wavelength $\sim q^{-1}_\text{sp}\equiv \sqrt{-2\kappa/\partial_{\phi}\tilde{\mu}_h}$ that coarsen over time, as established previously~\cite{Takatori2014,Stenhammar2013}.

How do the features of MIPS change upon the introduction of chemotaxis ($\text{Pe}_\text{C}>0$)? Given a constant and uniform $S$, the homogeneous state is now described by spatially-uniform ABP and chemoattractant profiles, $\phi(\textbf{x})=\phi_0$ and $\tilde{c}(\textbf{x})=\tilde{c}_0$, where $\tilde{c}_0$ is given by the steady-state solution to Eq.~\eqref{eqn::dcdt}, $ \tilde{c}_0=S/(k\phi_0)$. By perturbing this steady state with small-amplitude fluctuations $\delta \phi = \delta \hat{\phi} e^{i \mathbf{q}\cdot\mathbf{x}+\omega t}$ and $\delta \tilde{c} = \delta \hat{c} e^{i \mathbf{q}\cdot\mathbf{x}+\omega t}$ of spatial wavevector $\textbf{q}$ and growth rate $\omega$, we obtain the dispersion relation $\omega(q)$, given by Eq.~\eqref{eqn::omega} in ~[SI], where $q=|\mathbf{q}|$ is the wavenumber of a given mode. The homogeneous state is linearly stable if $\text{Re}~\omega<0$, which is always true when $\partial_\phi\tilde{\mu}_h>0$. We therefore focus our subsequent analysis on the spinodal region of non-chemotactic MIPS where $\partial_\phi\tilde{\mu}_h<0$, and nondimensionalize $\mathbf{q}$ and $\omega$ by the characteristic non-chemotactic MIPS quantities $q_\text{sp}$ and $\omega_\text{sp} \equiv \omega(q_\text{sp};\text{Pe}_\text{C}=0)$. As detailed in Sec.~\ref{sec::SI_LSA} in~[SI], the dispersion relation for chemotactic MIPS [Eq.~\eqref{eqn::nondim_eig_sln}] solely depends on three dimensionless parameters:
\begin{itemize}[noitemsep,nolistsep]
\item $\alpha \equiv - M_0 \phi_0 \partial_\phi\tilde{\mu}_h / D_c$, which compares the effective collective ABP diffusivity $- M_0 \phi_0 \partial_\phi\tilde{\mu}_h$ to that of the chemoattractant,
\item The Damk{\"o}hler number $\text{Da} \equiv k \phi_0 / (2D_c q_\text{sp}^2) = -\kappa k \phi_0 / (D_c \partial_\phi\tilde{\mu}_h)$, which compares the rates of chemoattractant uptake and diffusion over the characteristic length scale $q^{-1}_\text{sp}/\sqrt{2}$, and
\item The reduced chemotactic P\'{e}clet number $\text{Pe}_\text{C}' \equiv \chi_0 \tilde{c}_0 / (-M_0 \phi_0 \partial_\phi\tilde{\mu}_h)$.
\end{itemize}
Because the MIPS phase diagram is conventionally parameterized by $\phi_0$ and $\text{Pe}_\text{R}$, which together set $\partial_\phi\tilde{\mu}_h$ (Eq.~\eqref{eqn::dmu_h_dphi} in~[SI]), we also define versions of the three dimensionless parameters that are independent of these variables: $\alpha_0 \equiv M_0 / D_c$, $\text{Da}_0 \equiv \kappa k / D_c$, and $\text{Pe}_\text{C}$ given earlier, such that $\alpha = -\alpha_0\phi_0 \partial_\phi\tilde{\mu}_h$, $\text{Da} = -\text{Da}_0 \phi_0 /\partial_\phi\tilde{\mu}_h$, and $\text{Pe}_\text{C}'= -\text{Pe}_\text{C} \cdot S /( k \phi_0^2 \partial_\phi\tilde{\mu}_h)$. Furthermore, because the proportionality between $\text{Pe}_\text{C}'$ and $\text{Pe}_\text{C}$ is scaled by $S /k$, without loss of generality, we fix the chemoattractant supply rate $S/k=1$. Chemotactic MIPS is then parameterized by a total of five governing parameters: $\{\phi_0,\text{Pe}_\text{R},\alpha_0,\text{Da}_0,\text{Pe}_\text{C}\}$, as summarized in Table~\ref{table::variables} in~[SI]. Thus, to examine how chemotaxis influences MIPS, we first examine how the conventional $\phi_0-\text{Pe}_\text{R}$ phase diagram of MIPS changes upon varying $\alpha_0$, $\text{Da}_0$, and $\text{Pe}_\text{C}$.

As detailed in Sec.~\ref{sec::SI_stability_condition} in~[SI], our first main result from the linear stability analysis is that phase separation is suppressed by chemotaxis, but only when two criteria are \emph{simultaneously} satisfied: (1) $\text{Pe}_\text{C}' \geq \text{Pe}_{\text{C,crit}}'$, and (2) $\alpha \leq \alpha_\text{crit}$, where $\text{Pe}_{\text{C,crit}}' =  (1+ \text{min}\{\text{Da},1\})^2/(4\cdot \text{min}\{\text{Da},1\})$ and $\alpha_\text{crit} = 1+2\cdot\text{Da}+2\sqrt{\text{Da}(1+\text{Da})}$. We therefore designate the limits given by $\text{Pe}_\text{C}'=\text{Pe}'_{\text{C,crit}}$ and $\alpha=\alpha_\text{crit}$ as ``Boundary 1'' and ``Boundary 2''---shown in the $\text{Pe}_\text{R}-\phi_0$ phase diagrams (Fig.~\ref{fig::ABP}) by the solid and red dotted curves, respectively. Boundary 1 is colored by the different values of $\text{Pe}_\text{C}$. Boundary 2 does not depend on $\text{Pe}_\text{C}$. Criteria (1) and (2) correspond to the regions above Boundaries 1 and 2, respectively; hence, the region above \emph{both} Boundaries represents the stable regime in which the ABPs are in the homogeneous state, while conversely, the region below \emph{either} Boundary 1 \emph{or} 2 represents the unstable regime in which the ABPs phase separate. 

As a starting example, we examine the ABP phase diagram for $\text{Da}_0=0.2$ and $\alpha_0=1$, shown in Fig.~\ref{fig::ABP}(a). In this case, Boundary 2 is below the horizontal axis; hence, the system is linearly stable above Boundary 1 and unstable below it. Boundary 1 shifts to lower $\text{Pe}_\text{R}$ and a narrower range of $\phi_0$ with increasing $\text{Pe}_\text{C}$. That is, the region of instability shrinks, and phase separation is suppressed, when chemotaxis is stronger. Numerical simulations at $\text{Pe}_\text{C}=1$ confirm this linear stability result: ABPs are in the homogeneous state above Boundary 1, while phase separation occurs below it, as shown in Fig.~\ref{fig::ABP}(b). Intriguingly, the features of this phase separation appear to be fundamentally distinct from the spinodal decomposition observed in conventional non-chemotactic MIPS. For example, as shown in Movie~S2, ABPs phase separate into finite-sized domains that remain stationary, and do not subsequently coarsen---unlike in conventional MIPS.

Next, by increasing $\alpha_0$ to $4$, Boundary 1 remains unaltered, but Boundary 2 shifts downward,  as shown in Fig.~\ref{fig::ABP}(c). As a result, for the case of $\text{Pe}_\text{C}=1$, Boundary 2 rises above Boundary 1, which is omitted since Boundary 2 now corresponds to the limit of stability, as confirmed by numerical simulations shown in Fig.~\ref{fig::ABP}(d). As shown in Movie~S3, ABPs phase separate into finite-sized domains and bands that form traveling waves, a feature that is fundamentally distinct both from conventional MIPS and Fig. \ref{fig::ABP}(b). 

Finally, to highlight yet another distinct form of phase separation, we then increase both $\alpha_0$ and $\text{Da}_0$ in Fig.~\ref{fig::ABP}(e), where Boundary 1 shifts downward while Boundary 2 shifts upward, part of which becomes the limit of stability for $\text{Pe}_\text{C}=0.35$, confirmed by simulations in Fig.~\ref{fig::ABP}(f). Strikingly, we find that throughout the unstable region, the patterns vary from traveling bands that are extended (shaded green + hashed region) or less extended (unshaded + hashed region) to domains that stretch, rotate, and translate (unshaded region below the green dashed curve), as shown in Movie S4.

\begin{figure*}
  \includegraphics[width=0.9\textwidth]{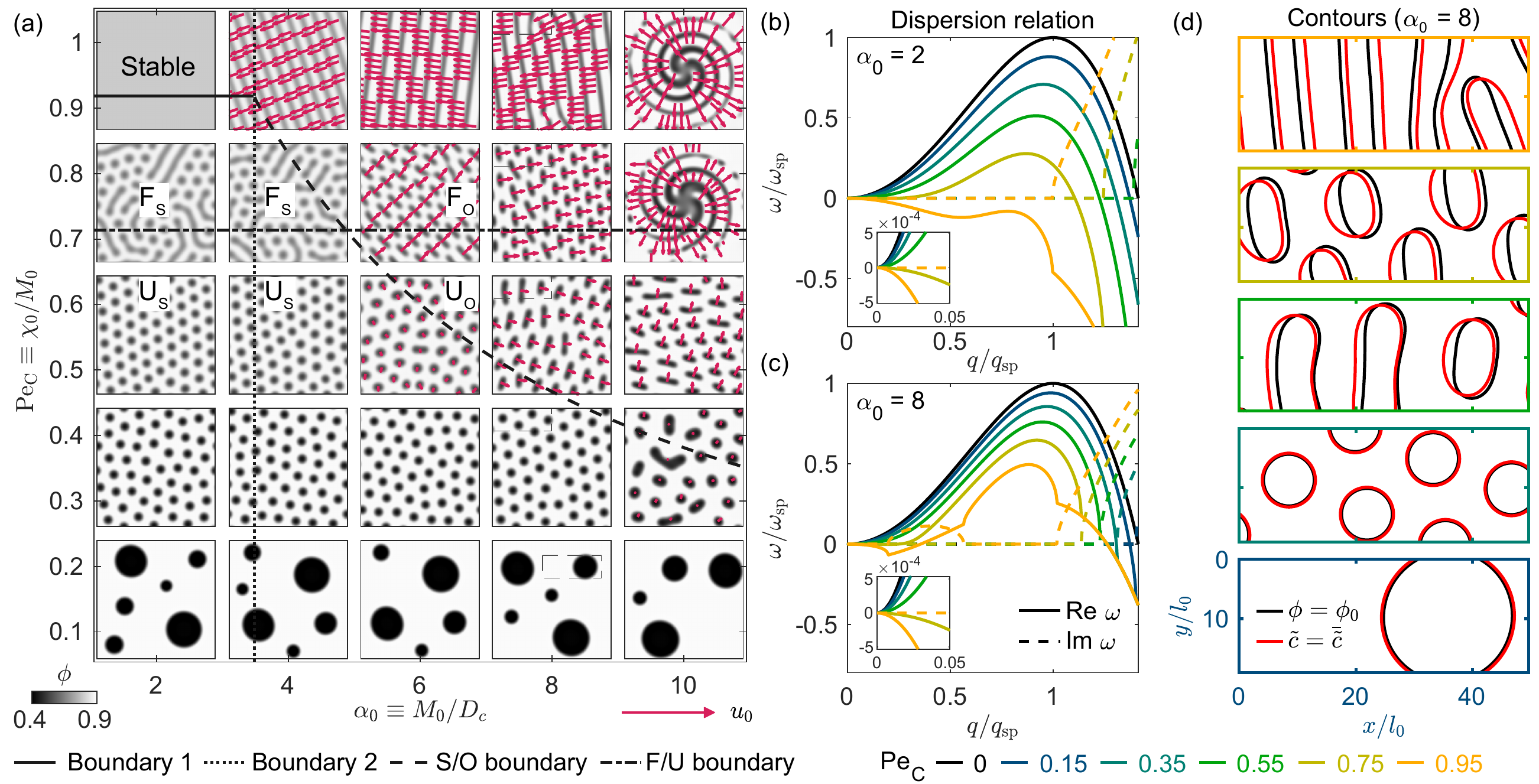}
  \caption{Chemotaxis arrests phase separation and generates dynamic instabilities. (a) Phase diagram parameterized instead by $\alpha_0$ and $\text{Pe}_\text{C}$, holding $\phi_0=0.8$, $\text{Pe}_\text{R}=10^{-3}$, and $\text{Da}_0=0.5$ fixed. Different instability types and the boundaries between them, as predicted by our linear stability analysis, are indicated using the same labels as in Fig.~\ref{fig::ABP}. These predictions are again corroborated by numerical simulations (Movie S7), snapshots of which are shown. Arrows show the local velocity field $\mathbf{u}$, with the scale indicated by the characteristic velocity $u_0 \equiv M_0/\sqrt{\kappa} \sim U_0$; velocities for which $|\mathbf{u}|<0.005u_0$ are not shown. (b-c) Dispersion relations $\omega(q)$ corresponding to $\alpha_0=2$ and $\alpha_0=8$ in (a); solid (dashed) lines show the Real (Imaginary) components. Insets zoom in on long wavelengths. (d) Magnified views of the contours of $\phi=\phi_0$ and $\tilde{c}=\bar{\tilde{c}}$ (the spatial average of $\tilde{c}$) for the small regions indicated by the dashed rectangles in the snapshots of (a) at $\alpha_0=8$. Different colors in (b-d) show the different values of $\text{Pe}_\text{C}$ corresponding to the simulations shown in (a).}
  \label{fig::chi_alpha0}
\end{figure*}

Taken altogether, these results demonstrate that MIPS is suppressed when (1) the strength of chemotaxis, as quantified by $\text{Pe}_\text{C}$, \emph{and} (2) chemoattractant diffusivity relative to that of the ABPs, as quantified by $\alpha_{0}^{-1}$, are sufficiently high. Moreover, our simulations reveal that the features of phase separation are dramatically altered by chemotaxis---with separated domains that initially can either be finite-sized or unbounded in space, and can either be stationary or exhibit complex oscillatory dynamics in time, depending on the values of  $\{\phi_0,\text{Pe}_\text{R},\alpha_0,\text{Da}_0,\text{Pe}_\text{C}\}$. We summarize these results in the $\alpha_0-\text{Pe}_\text{C}$ phase diagram shown in Fig.~\ref{fig::chi_alpha0}(a), holding $\phi_0$, $\text{Pe}_\text{R}$, and $\text{Da}_0$ fixed, and show the region of stability (which lies above Boundary 1 and to the left of Boundary 2 in the $\alpha_0-\text{Pe}_\text{C}$ plane shown) and snapshots of these different types of instability (animated in Movie S7) that we now seek to categorize.

\textbf{\emph{Chemotaxis arrests phase separation.}} We first classify the instabilities by their distinct spatial characteristics. In particular, depending on the range of initially-unstable wavenumbers $q_-<q<q_+$ in the dispersion relation $\omega(q)$ (Eq.~\eqref{eqn::omega} in~[SI]) derived using our linear stability analysis, we differentiate instabilities as being either finite-wavelength~(F) when the unstable modes are spatially bounded ($q_->0$), and therefore phase-separated domains do not coarsen, or unbounded~(U) when the unstable modes can instead extend indefinitely in space ($q_-=0$)~\cite{Worlitzer2021}.
While conventional MIPS is a Type~U instability~\cite{Takatori2014,Stenhammar2013,Cross1993}, our second main result is that chemotaxis can give rise to Type~F instabilities as well---as shown by the domains that do not coarsen in e.g., Movies S2--S3 noted earlier. Comparing the ABP (Movies S2--3) and chemoattractant (Movies S5--6) profiles reveals the underlying reason: ABPs in an extended, dense domain collectively establish a strong local chemoattractant gradient through uptake---which in turn causes them to bias their motion up the gradient and disperse away, arresting phase separation.

This behavior is also reflected in the simulations shown in Fig.~\ref{fig::chi_alpha0}(a) and Movie S7. For the example of $\alpha_0=2$ (left of Boundary 2), as $\text{Pe}_\text{C}$ increases, the coarsening slows and eventually becomes arrested (Sec.~\ref{sec::SI_coarsen} in~[SI]), forming finite-sized domains and stripes---ultimately reaching the homogeneous state at the largest $\text{Pe}_\text{C}$ above Boundary 1. Examining the dispersion relations in Fig.~\ref{fig::chi_alpha0}(b) and the inset corroborates this observation. At low non-zero $\text{Pe}_\text{C}$, the unstable modes extend to $q_{-}=0$ (blue to green curves), indicating a Type U instability. By contrast, for the larger $\text{Pe}_\text{C}=0.76$, $q_{-}>0$ (chartreuse curve), indicating a Type F instability.

Indeed, determining $q_-$ directly from the dispersion relation as described in Sec.~\ref{sec::SI_FU_instability} in~[SI] yields the quantitative criterion that Type F is given by $\text{Pe}_\text{C}'>1$ (shaded regions in Fig.~\ref{fig::ABP}), while Type U is given by $\text{Pe}_\text{C}'<1$ (unshaded). The boundary between the two, given by $\text{Pe}_\text{C}'=1$ (Eq.~\ref{eqn::I_II_boundary} in~[SI]), is represented by the dash-dotted curves in Figs.~\ref{fig::ABP} and \ref{fig::chi_alpha0}(a). In all cases, our predictions for the Type F/U boundary agree well with the simulations, as detailed in Sec.~\ref{sec::SI_dispersion} in~[SI]---
thereby providing a quantitative description of how chemotaxis can arrest MIPS.

\textbf{\emph{Chemotaxis engenders complex oscillatory dynamics.}} Following Cross and Hohenberg~\cite{Cross1993}, we further classify the instabilities by their distinct temporal characteristics -- ``Stationary'' (S) if all unstable modes are non-oscillatory with $\text{Im}~\omega=0$, or ``Oscillatory'' (O) if there exist unstable and oscillatory modes with $\text{Re}~\omega(q)>0$ and $\text{Im}~\omega(q) \neq 0$. While conventional MIPS is a Type S instability, our third main result is that chemotaxis can give rise to Type O instabilities as well---e.g., Movies S3-4 noted earlier. This behavior is also reflected in Fig.~\ref{fig::chi_alpha0}(a) and Movie S7. As shown in Fig.~\ref{fig::chi_alpha0}(c) for the example of $\alpha_0=8$, at low $\text{Pe}_\text{C}$ (blue and cyan curves), all unstable modes (with $\text{Re}~\omega>0$) are stationary (having $\text{Im}~\omega= 0$), indicating a Type S instability; by contrast, at higher $\text{Pe}_\text{C}$ (green to orange curves), some unstable modes have $\text{Im}~\omega\neq 0$, indicating a Type O instability. As a result, in this regime, the phase-separated domains continually move in complex ways---e.g., stretching, rotating, and translating---as indicated by the arrows in Fig.~\ref{fig::chi_alpha0}(a) showing the local velocity field $\mathbf{u}$.

Why do these complex dynamics emerge for sufficiently strong chemotaxis (large $\text{Pe}_\text{C}$) and slow chemoattractant diffusion (large $\alpha_0$)? Comparing the ABP and chemoattractant profiles, $\phi(\textbf{x})$ and $\tilde{c}(\textbf{x})$ respectively, again sheds light on the underlying physics. Fig.~\ref{fig::chi_alpha0}(d) shows the illustrative case of $\alpha_0=8$ for the five different $\text{Pe}_\text{C}$ shown in (a). For the lowest two $\text{Pe}_\text{C}$, chemotaxis is weak, enabling $\tilde{c}(\textbf{x})$ to equilibrate in response to changes in $\phi(\textbf{x})$. Consequently, the phase-separated patterns remain stationary, reflective of a Type S instability. For larger $\text{Pe}_\text{C}$, however, chemotaxis proceeds more rapidly and the diffusing chemoattractant cannot equilibrate fast enough. As a result, variations in $\tilde{c}(\textbf{x})$ lag behind $\phi(\textbf{x})$, driving directed large-scale motion of the phase-separated domains, reflective of a Type O instability. Intriguingly, a similar mechanism has been proposed to explain the spontaneous autophoresis of chemically-active particles~\cite{Michelin2013,Michelin2014}.

The dispersion relation again yields a quantitative criterion for the Type O instability, shown as the hashed regions in Fig.~\ref{fig::ABP}. The Type S/O boundary given by Eq.~\eqref{eqn::critical_condition_oscillatory} in~[SI] is represented using the dashed curves in Figs.~\ref{fig::ABP}(e)-(f) and \ref{fig::chi_alpha0}(a); in Fig.~\ref{fig::ABP}(c)-(d), this Boundary coincides with Boundary 2. We again observe good agreement betweeb the predicted Type S/O boundary and the numerical simulations \footnote{We note, however, that below the S/O boundary shown by the dashed curve in Fig.~\ref{fig::ABP}(e)--(f), the simulations still show some initial non-stationary behavior---reflecting the limitation of our linear stability analysis, which is strictly only applicable to conditions close to the initial homogeneous state.}. Thus, our analysis provides a quantitative explanation of how the interplay between chemotaxis and chemoattractant diffusion can generate more complex phase separation dynamics than in conventional MIPS.

\textbf{\emph{Discussion.}}
Motivated by the prevalence of chemotaxis in active systems, we have developed an illustrative model of chemotactic MIPS. We find that chemotaxis strongly competes with MIPS---in some cases, arresting or completely suppressing phase separation, or in other cases, generating fundamentally new dynamic instabilities that share features with other pattern-forming systems, but arise due to completely different physics~\cite{Cross1993,Kondo2010,Bar2020,Zwicker2015,Zwicker2017,Menzel2013,Ziepke2022,Saha2020,You2020Marchetti,vanderKolk2022,Matas-Navarro2014,Navarro2015,Yin2022,Bazant2017,Adkins2022,Tayar2022,Caballero2022}. 
This work thus helps to reveal and clarify the rich new physics underlying active systems that perform chemotaxis, ranging from enzymes at the subcellular scale to collectives of living cells and chemically-active colloids and beyond. 

Our work also provides quantitative guidelines to rationalize existing observations and guide new experiments to search for the fascinating behaviors predicted here. For example, simple estimates based on our findings (Sec.~\ref{sec::SI_application} in~[SI]) suggest that chemotaxis may help suspensions of motile microorganisms overcome MIPS and remain in the homogeneous state under nutrient-replete conditions. When starved, however, our analysis suggests that such suspensions will separate into dense communities that may confer functional benefits---as has indeed been observed in many experiments~\cite{Liu2019,Budrene1991}. We also expect that the different instabilities described here could be explored using synthetic forms of active matter with tunable velocities and chemical dynamics, as detailed further in Sec.~\ref{sec::SI_application} in~[SI]. More broadly, while we focused on biased motion up a chemoattractant gradient as an illustrative example, our theoretical framework also provides a foundation to describe the influence of chemorepulsion, as well as other forms of taxis---e.g., durotaxis, electrotaxis, and phototaxis~\cite{Roca-Cusachs2013,Shellard2020,SenGupta2021,Sunyer2016,Alert2019a,Cohen2014,Mijalkov2016,Palagi2018}---on MIPS. 

\begin{acknowledgments}
We acknowledge support from NSF Grants CBET-1941716 and DMR-2011750, the Pew Biomedical Scholars Program, and a Princeton Bioengineering Initiative (PBI$^2$) Postdoctoral Fellowship.
\end{acknowledgments}

\newpage
\setcounter{figure}{0}

\makeatletter 
\renewcommand{\thefigure}{S\@arabic\c@figure}
\makeatother

\section*{{Supplementary Information}}

\section{Thermodynamics of non-chemotactic ABPs}
\label{sec::SI_ABP}
As derived in~\cite{Takatori2015}, the non-dimensional active pressure generated by ABPs in 2D is
\begin{equation} \label{eqn::Pi_2D}
  \frac{\Pi}{n \zeta U_0^2 \tau_R / 2} = 1 - \phi - 0.2 \phi^2 + \frac{4}{\pi} \phi \text{Pe}_\text{R} \left( 1 - \frac{\phi}{\phi_m} \right)^{-1},
\end{equation}
where $n$ is the particle number density, $\phi$ is the area fraction $\phi = n v_0$, where $v_0=\pi a^2$ is the area taken up by each particle, and $\phi_m = 0.9$ is the maximum area fraction ($0\leq \phi < \phi_m$). This pressure is also related to a nonequilibrium Helmholtz free energy per volume $f$,
\begin{equation}
  \label{eqn::Pi_to_f}
  \Pi = \phi^2 \pderiv{}{\phi}\left(\frac{f}{\phi}\right) = f'\phi - f.
\end{equation}
Combining Eqs.~\eqref{eqn::Pi_2D}- \eqref{eqn::Pi_to_f} then yields
\begin{multline}
  \frac{f}{\zeta U_0^2 \tau_R/2} = \frac{\phi}{v_0} \bigg[ \ln{\phi} - \phi - 0.1\phi^2 \big. \\ \left. - \frac{4}{\pi} \text{Pe}_\text{R}\cdot \phi_m \ln{\left(1 - \frac{\phi}{\phi_m}\right)} \right].
\end{multline}
Given this Helmholtz free energy, one can further define the bulk chemical potential, which we use in the calculations described in the main text:
$\mu_h \equiv \partial f/ \partial n = \partial (v_0 f)/\partial \phi$. 
As explained in the main text, we define a nondimensionalized version of it as $\tilde{\mu}_h \equiv \mu_h / (\zeta U_0^2 \tau_R/2)$. This definition yields
\begin{multline}
  \label{eqn::mu_h}
  \tilde{\mu}_h = \ln{\phi}+1 - 2\phi - 0.3\phi^2 \\ - \frac{4}{\pi}\text{Pe}_\text{R}\cdot \phi_m \left[ \ln{\left( 1- \frac{\phi}{\phi_m}\right)} - \frac{\phi}{\phi_m - \phi} \right].
\end{multline}
When analyzing the linear stability in Sec.~\ref{sec::SI_LSA}, we often need to evaluate the derivative of the non-dimensional chemical potential with respect to $\phi$,
\begin{equation}
  \label{eqn::dmu_h_dphi}
  \partial_\phi \tilde{\mu}_h = \frac{1}{\phi} - 2 - 0.6 \phi - \frac{4}{\pi} \text{Pe}_\text{R} \cdot\frac{\phi_m (\phi - 2\phi_m)}{(\phi-\phi_m)^2}.
\end{equation}
It is useful to note that $-\partial_\phi\tilde{\mu}_h$ has an upper bound:
\begin{eqnarray} \label{eqn::dmu_upper_bound}
  \sup_{\phi,\text{Pe}_\text{R}} \left({-\partial_\phi \tilde{\mu}_h}\right) &=& \lim_{\phi\to\phi_m,\text{Pe}_\text{R}\to0}{-\partial_\phi \tilde{\mu}_h} \nonumber \\
  &=& -\frac{1}{\phi_m}+2 + 0.6\phi_m \approx 1.43.
\end{eqnarray}

Consistent with the classical Cahn-Hilliard theory of phase separation, the free energy can be extended to penalize a sharp interface~\cite{Cates2013,Stenhammar2013}. The total free energy in a spatial field is
\begin{equation}
  F = \int{ \left( f + \frac{1}{2} \frac{\zeta U_0^2 \tau_R}{2v_0} \kappa \|\nabla \phi \|^2 \right)  d\mathbf{x}},
\end{equation}
from which the overall chemical potential can be defined variationally by $\mu \equiv \delta F / \delta n = v_0 \delta F / \delta \phi $; here, $\kappa=l_0^2$ as noted in the main text.
Again, we define a normalized version of this overall chemical potential $\tilde{\mu} \equiv \mu / (\zeta U_0^2 \tau_R/2)$. Therefore,
\begin{equation}
  \tilde{\mu} = \tilde{\mu}_h - \kappa \nabla^2 \phi,
\end{equation}

Using these thermodynamic rules, we next describe the phase dynamics following Ref.~\cite{Takatori2015}. 
The particle volume fraction satisfies the conservation equation:
\begin{equation}
  \pderiv{\phi}{t} =  \nabla \cdot \left(\frac{\phi}{\zeta}  \nabla \mu\right) = \nabla \cdot \bigg(M_0 \phi \nabla \big( \tilde{\mu}_h(\phi,\text{Pe}_\text{R}) - \tilde{\kappa} \nabla^2 \phi \big)\bigg),
\end{equation}
where $M_0 = U_0^2 \tau_R/2$. For convenience of notation, we define the collective diffusivity $M(\phi) \equiv M_0\phi$. This expression thereby yields the part of Eq.~(1) of the main text that reflects active Brownian motion.

\section{Linear stability analysis}
\label{sec::SI_LSA}
\subsection{Dispersion relation}
\label{sec::SI_dispersion}
In this section, we study the linear stability of the governing equations, Eqs.~(1)-(3) of the main text. For generality, here we do not assume any particular functional form for the chemotactic sensing function $f(\tilde{c})$ or chemoattractant uptake rate $g(\tilde{c})$.
We perturb the homogeneous steady state $\phi(\mathbf{x})=\phi_0$ and $\tilde{c}(\mathbf{x})=\tilde{c}_0=g^{-1}(S k^{-1} \phi_0^{-1})$ with small amplitude perturbations $\delta \phi = \delta \hat{\phi} e^{i \mathbf{q}\cdot\mathbf{x} + \omega t}$, and $\delta \tilde{c} = \delta \hat{c} e^{i \mathbf{q}\cdot\mathbf{x} + \omega t}$.
Linearizing Eqs.~(1)-(3) and substituting $\delta \phi$ and $\delta \tilde{c}$ yields
\begin{align}
  \omega \delta \hat{\phi} &= -M(\phi_0) q^2 (\partial_\phi \tilde{\mu}'_h(\phi_0) + \kappa q^2) \delta \hat{\phi} + \chi_0 \phi_0 q^2 f'(\tilde{c}_0) \delta \hat{c}, \\
  \omega \delta \hat{c} &= - D_c q^2 \delta \hat{c} - k( g(\tilde{c}_0) \delta \hat{\phi} + \phi_0 g'(\tilde{c}_0) \delta \hat{c}),
\end{align}
where $q = |\mathbf{q}|$. For simplicity of notation, in the following text, the arguments $\phi_0$ and $\tilde{c}_0$ in $\partial_\phi \tilde{\mu}'_h(\phi_0)$, $g(\tilde{c}_0)$, $g'(\tilde{c}_0)$, and $f'(\tilde{c}_0)$ are omitted. The eigenvalue $\omega$ satisfies
\begin{equation}
  \label{eqn::LSA_quadratic_form}
  \omega^2 + (\mathcal{M}+\mathcal{D}) \omega + \mathcal{M}\mathcal{D}+\mathcal{X} = 0.
\end{equation}
The solution to $\omega$ is
\begin{equation} \label{eqn::omega}
  \omega_\pm = \frac{1}{2}\left( -(\mathcal{M}+\mathcal{D}) \pm \sqrt{(\mathcal{M}-\mathcal{D})^2 -4\mathcal{X}} \right),
\end{equation}
where
\begin{align}
  \begin{split}
    \label{eqn::LSA_three_terms}
    \mathcal{M} &\equiv M q^2 (\partial_\phi \tilde{\mu}_h + \kappa q^2), \\
    \mathcal{D} &\equiv D_c q^2 + k \phi_0 g', \\
    \mathcal{X} &\equiv k \chi_0 \phi_0 f' g q^2.
  \end{split}
\end{align}
In conventional non-chemotactic MIPS ($\chi_0=0$), the two eigenvalues are $-\mathcal{M}$ and $-\mathcal{D}$, respectively. Because $\mathcal{D} \geq 0$, the stability is determined by $\mathcal{M}$. When $\partial_\phi \tilde{\mu}_h<0$, or in the spinodal region as defined in the main text, $\omega$ can be positive in a range of wavenumber $q$, and the most unstable wavenumber that corresponds to maximum instability growth rate $\omega$ is $q^{-1}_\text{sp}\equiv \sqrt{-2\kappa/\partial_{\phi}\tilde{\mu}_h}$.
Because of Eq.~\eqref{eqn::dmu_upper_bound}, $q_\text{sp}^{-1} \gtrsim l_0$.
By nondimensionalizing wavenumber with the characteristic length  scale of spinodal decomposition,
\begin{equation}
  \tilde{q} \equiv \sqrt{-\frac{\kappa}{\partial_\phi \tilde{\mu}_h}} q = \frac{q}{\sqrt{2} q_\text{sp}},
\end{equation}
and nondimensionalizing rate with the characteristic growth rate of non-chemotactic spinodal decomposition~$4\omega_\text{sp}= 4\omega(q_\text{sp},\text{Pe}_\text{C}=0)=M\left(\partial_\phi\tilde{\mu}_h\right)^2/\kappa$, $\tilde{\omega}\equiv\omega/(4\omega_\text{sp})$,
$\tilde{\mathcal{M}}\equiv\mathcal{M}/(4\omega_\text{sp})$, $\tilde{\mathcal{D}}\equiv\mathcal{D}/(4\omega_\text{sp})$, $\tilde{\mathcal{X}}\equiv\mathcal{X}/(4\omega_\text{sp})^2$, we obtain the nondimensionalized equation for the eigenvalues
\begin{equation}
  \label{eqn::nondim_eig}
  \tilde{\omega}^2 + (\tilde{\mathcal{M}}+\tilde{\mathcal{D}}) \tilde{\omega} + \tilde{\mathcal{M}}\tilde{\mathcal{D}}+\tilde{\mathcal{X}} = 0.
\end{equation}
The solution is then
\begin{equation}
  \label{eqn::nondim_eig_sln}
  \tilde{\omega}_\pm = \frac{1}{2}\left( -(\tilde{\mathcal{M}}+\tilde{\mathcal{D}}) \pm \sqrt{(\tilde{\mathcal{M}}-\tilde{\mathcal{D}})^2 -4\tilde{\mathcal{X}}} \right),
\end{equation}
where
\begin{align}
  \begin{split}
    \tilde{\mathcal{M}} &= \tilde{q}^2 \left( -1 + \tilde{q}^2 \right), \\
    \tilde{\mathcal{D}} &= \frac{1}{\alpha} \left( \tilde{q}^2 + \text{Da} \right),  \\
    \tilde{\mathcal{X}} &= \frac{\text{Da}}{\alpha} \text{Pe}_\text{C}' \tilde{q}^2,
  \end{split}
\end{align}
and the dimensionless parameters are
\begin{align} \label{eqn::dimensionless_parameters}
  \begin{split}
    \alpha &= -\frac{M\partial_\phi\tilde{\mu}_h}{D_c}, \\
    \text{Da} &= -\frac{\kappa k\phi_0 g'}{D_c \partial_\phi \tilde{\mu}_h}, \\
    \text{Pe}_\text{C}' &= -\frac{\chi_0}{M\partial_\phi \tilde{\mu}_h} \frac{f' g}{g'}.
  \end{split}
\end{align}
We restrict our discussion below to $\partial_\phi \tilde{\mu}_h<0$ (in the spinodal region), $\alpha>0$, $\text{Da}>0$, and $\text{Pe}_\text{C}' \geq 0$.

\subsection{Stability condition}
\label{sec::SI_stability_condition}
When the discriminant of the quadratic equation Eq.~\eqref{eqn::nondim_eig} is positive, i.e., $\Delta \equiv (\tilde{\mathcal{M}}-\tilde{\mathcal{D}})^2 -4\tilde{\mathcal{X}} > 0$, it can be seen from Eq.~\eqref{eqn::nondim_eig_sln} that $\tilde{\omega}_+$ decreases with increasing $\text{Pe}_\text{C}'$. In other words, chemotaxis has a stabilizing effect. Therefore, next, we derive the condition under which the system is stable, that is, $\text{Re}~\tilde{\omega}_\pm(\tilde{q}) \leq 0$ for all $\tilde{q}$. This condition is equivalent to (1) $I_2 \equiv \tilde{\omega}_+ \tilde{\omega}_- = \tilde{\mathcal{M}}\tilde{\mathcal{D}}+\tilde{\mathcal{X}} \geq 0$, \textit{and} (2) $I_1 \equiv \tilde{\omega}_+ + \tilde{\omega}_- = - (\tilde{\mathcal{M}}+\tilde{\mathcal{D}})\leq 0 $ for all $\tilde{q}$.

Criterion (1) ($I_2 \geq 0$) can be achieved with sufficiently large $\text{Pe}_\text{C}'$: since
\begin{equation}
  I_2 = \frac{\tilde{q}^2}{\alpha}\left( (\tilde{q}^2-1)(\tilde{q}^2+\text{Da}) + \text{Da} \text{Pe}_\text{C}' \right),
\end{equation}
$I_2 \geq 0$ for all $\tilde{q}$ is equivalent to $\min_{\tilde{q}}{\alpha\tilde{q}^{-2}I_2} \geq 0$.
When $\text{Da} \leq 1$, the minimum is obtained at $\tilde{q}=0$, and $\min_{\tilde{q}}{\alpha\tilde{q}^{-2}I_2}=\text{Da}(\text{Pe}_\text{C}'-1)$; hence, criterion (1) is equivalent to $\text{Pe}_\text{C}'>1$. When $\text{Da}>1$, the minimum is obtained at $\tilde{q}^2 = (1-\text{Da})/2$,
and
\begin{equation}
  \label{eqn::min_Re_omega}
  \min_{\tilde{q}}{\alpha\tilde{q}^{-2}I_2} = -\frac{(1+\text{Da})^2}{4} + \text{Da}\text{Pe}_\text{C}'.
\end{equation}
In this case criterion (1) is equivalent to $\text{Pe}_\text{C}'>(1+\text{Da})^2/(4\text{Da})$.
Therefore, we can summarize criterion (1) in a more compact form as
\begin{equation} \label{eqn::critical_condition_Pe}
  \text{Pe}_\text{C}' > \text{Pe}_\text{C,crit}' = \frac{(1+ \text{min}\{\text{Da},1\})^2}{4 \text{min}\{\text{Da},1\}}.
\end{equation}
In other words, in order to suppress phase separation, chemotactic rate needs to be sufficiently fast.

As noted above, at the critical point of stability where $\max_{\tilde{q}}{I_2}=0$, the critical wavenumber is
\begin{equation}
  \tilde{q}^2_\text{crit,2} = \frac{1-\text{min}\{\text{Da},1\}}{2}.
\end{equation}
This result indicates that if criterion (2) is satisfied so that the stability of the system is solely determined by criterion (1), as the control parameter $\text{Pe}_\text{C}'$ varies near the critical condition of stability, the range of unstable wavelength can either be unbounded ($\tilde{q}$ near 0) if $\text{Da} \geq 1$ or finite ($\tilde{q}$ near $\sqrt{(1-\text{Da})/2}$) if $\text{Da} \leq 1$. The former belongs to type F instability while the latter belongs to type U instability according to Cross and Hohenberg's classification of dispersion relations~\cite{Cross1993}.

Having large $\text{Pe}_\text{C}'$ is a necessary but insufficient condition for the suppression of phase separation.
Another way to interpret criterion (2) ($I_1 \leq 0$) is that, when $\text{Pe}_\text{C}'$ is sufficiently large, $\Delta$ becomes negative, and $\text{Re}~\tilde{\omega}_\pm = -(\tilde{\mathcal{M}}+\tilde{\mathcal{D}})/2=I_1/2$. 
Hence, sufficiently large $\text{Pe}_\text{C}'$ can fully stabilize the system only when $I_1<0$.

Since
\begin{equation} \label{eqn::max_Re_omega_large_PeC}
  I_1 = -q^4 + \left(1-\frac{1}{\alpha}\right)\tilde{q}^2 - \frac{\text{Da}}{\alpha},
\end{equation}
when $\alpha \leq 1$, the maximum is obtained at $\tilde{q}=0$, and $\max_{\tilde{q}}{I_1}=-\text{Da}/\alpha<0$. When $\alpha>1$, the maximum is obtained at $\tilde{q}^2 = (\alpha-1)/2\alpha$,
and
\begin{equation}
  \label{eqn::max_Re_omega}
  \max_{\tilde{q}}{I_1} = \frac{1}{\alpha}\left(\frac{(\alpha-1)^2}{4\alpha} - \text{Da}\right).
\end{equation}
$I_1 \leq 0$ for all $\tilde{q}$ is equivalent to $\max_{\tilde{q}}{I_1}<0$, or equivalently $\alpha \leq 1$ or $\text{Da} \geq (\alpha-1)^2/4\alpha$. This condition can be written in a more compact form as shown in the main text,
\begin{equation}
  \alpha \leq \alpha_\text{crit} = 1+2\cdot\text{Da}+2\sqrt{\text{Da}(1+\text{Da})}.
\end{equation}
Or alternatively,
\begin{equation}
  \text{Da} \geq \text{Da}_\text{crit} = \frac{(1-\text{max}\{\alpha,1\})^2}{4\text{max}\{\alpha,1\}}.
\end{equation}
In other words, in order to suppress phase separation, chemoattractant diffusion or uptake rate needs to be sufficiently fast.

As noted above, at the critical condition of stability where $\max_{\tilde{q}}{I_1}=0$, the critical wavenumber is
\begin{equation}
  \tilde{q}^2_\text{crit,1} = \frac{1-\alpha^{-1}}{2}.
\end{equation}
This result indicates that if criterion (1) is satisfied so that the stability of the system is solely determined by criterion (2), as the control parameter $\alpha$ or $\text{Da}$ varies near the critical condition of stability, the range of unstable mode is finite and near $\tilde{q}_\text{crit,1}$. This belongs to type F instability according to Cross and Hohenberg's classification of dispersion relations~\cite{Cross1993}.

In summary, the stability criteria (1) and (2) are equivalent to $\text{Pe}_\text{C}' \geq \text{Pe}_{\text{C,crit}}'$ and $\alpha \leq \alpha_{\text{crit}}$ (or $\text{Da} \geq \text{Da}_\text{crit}$), indicating that MIPS can be suppressed with sufficiently fast chemotaxis and chemoattractant diffusion (or uptake rate).

\subsection{Finite and unbounded wavelength instabilities}
\label{sec::SI_FU_instability}
In the main text, we define finite and unbounded wavelength instabilities based on the range of unstable modes, which we here express in dimensionless form: $\tilde{q}_{u-}<\tilde{q}<\tilde{q}_{u+}$. When $\tilde{q}_{u-}=0$, the unstable wavelength extends all the way to infinity---we thus call this an unbounded instability (type U). Otherwise when $\tilde{q}_{u-}>0$, the range of unstable wavelengths is finite---we thus call this a finite wavelength instability (type F).

From Eq.~\eqref{eqn::nondim_eig_sln}, we see that $\tilde{\omega}_+(\tilde{q}=0)=0$ and $\tilde{\omega}'_+(\tilde{q}=0)=0$. Hence the sign of the second order derivative determines whether modes near $\tilde{q}=0$ are stable. At $\tilde{q}=0$,
\begin{equation}
  \tilde{\omega}''_+ = - \left. \frac{(\tilde{\mathcal{M}}\tilde{\mathcal{D}}+\tilde{\mathcal{X}})''}{\tilde{\mathcal{M}}+\tilde{\mathcal{D}}} \right|_{\tilde{q}=0} = 2 (1-\text{Pe}_\text{C}').
\end{equation}
When $\text{Pe}_\text{C}'<1$, $\tilde{\omega}''(\tilde{q}=0)>0$, we have $\tilde{q}_{u-}=0$, hence the system has 
an unbounded instability. Otherwise, when $\text{Pe}_\text{C}'>1$ and the system is in the unstable regime, it has a finite wavelength instability.
These results suggest that as the chemotactic rate increases, modes near zero wavenumber become stabilized---and thus, phase separated domains are less likely to coarsen since chemotaxis disperses the particles.

When criterion (2) described in Sec.~\ref{sec::SI_stability_condition} is satisfied, the dispersion relation can be classified by $\text{Pe}_\text{C}'$. 
If $\text{Da}<1$, the system has type U instability when $\text{Pe}_\text{C}'<1$, type F instability when $1<\text{Pe}_\text{C}'<\text{Pe}_{\text{C,crit}}'$, and is stable when $\text{Pe}_\text{C}'>\text{Pe}_{\text{C,crit}}'$. The transition from instability to stability by increasing chemotactic rate is of type F~\cite{Cross1993}. Hence, when chemoattractant uptake rate is slow such that $\text{Da}<1$, finite-sized domains can be observed near the boundary of stability.

If $\text{Da}>1$, the system has type U instability when $\text{Pe}_\text{C}'<1$, and is stable when $\text{Pe}_\text{C}'>1$.
The transition from instability to stability by increasing chemotactic rate is of type U~\cite{Cross1993}. Hence when chemoattractant uptake rate is fast such that $\text{Da}>1$, phase separated domains are more likely to coarsen near the boundary of stability.

The classification of type F/U instability also applies when criterion (2) is not satisfied, which we describe in the next section.

\subsection{Oscillatory instability condition}
\label{sec::oscillatory_condition}
In Sec.~\ref{sec::SI_stability_condition} we have shown that when criterion (2) is not satisfied, large $\text{Pe}_\text{C}'$ cannot suppress phase separation. Instead, at high enough $\text{Pe}_\text{C}'$, the discriminant becomes negative $\Delta<0$, which means that eigenvalues can have imaginary part ($\text{Im}~\tilde{\omega} \neq 0$). Therefore, next, we derive the condition for oscillatory instability---that there exists $\tilde{q}$ for which $\text{Re}~\tilde{\omega}>0$ and $\text{Im}~\tilde{\omega} \neq 0$, or equivalently $I_1>0$ and $\Delta<0$.

Since $I_1$ is a quadratic polynomial of $\tilde{q}^2$, $I_1>0$ can be obtained by finding the values of $\tilde{q}^2$ that correspond to the zeros of $I_1$:
\begin{equation}
  \tilde{q}_\pm^2 = \frac{\alpha-1 \pm \sqrt{(\alpha-1)^2-4\alpha \text{Da}}}{2\alpha}.
\end{equation}
In this section, we always require that $\alpha>1$ and $\text{Da}<(\alpha-1)^2/4\alpha$ (criterion (2) is not satisfied). This ensures that $\tilde{q}_\pm^2$ exist and are positive.
$I_1>0$ when $\tilde{q}_- < \tilde{q} < \tilde{q}_+$. Notice that $\tilde{q}_+ < (\alpha-1)/\alpha < 1$.

\begin{figure}
  \includegraphics[width=\columnwidth]{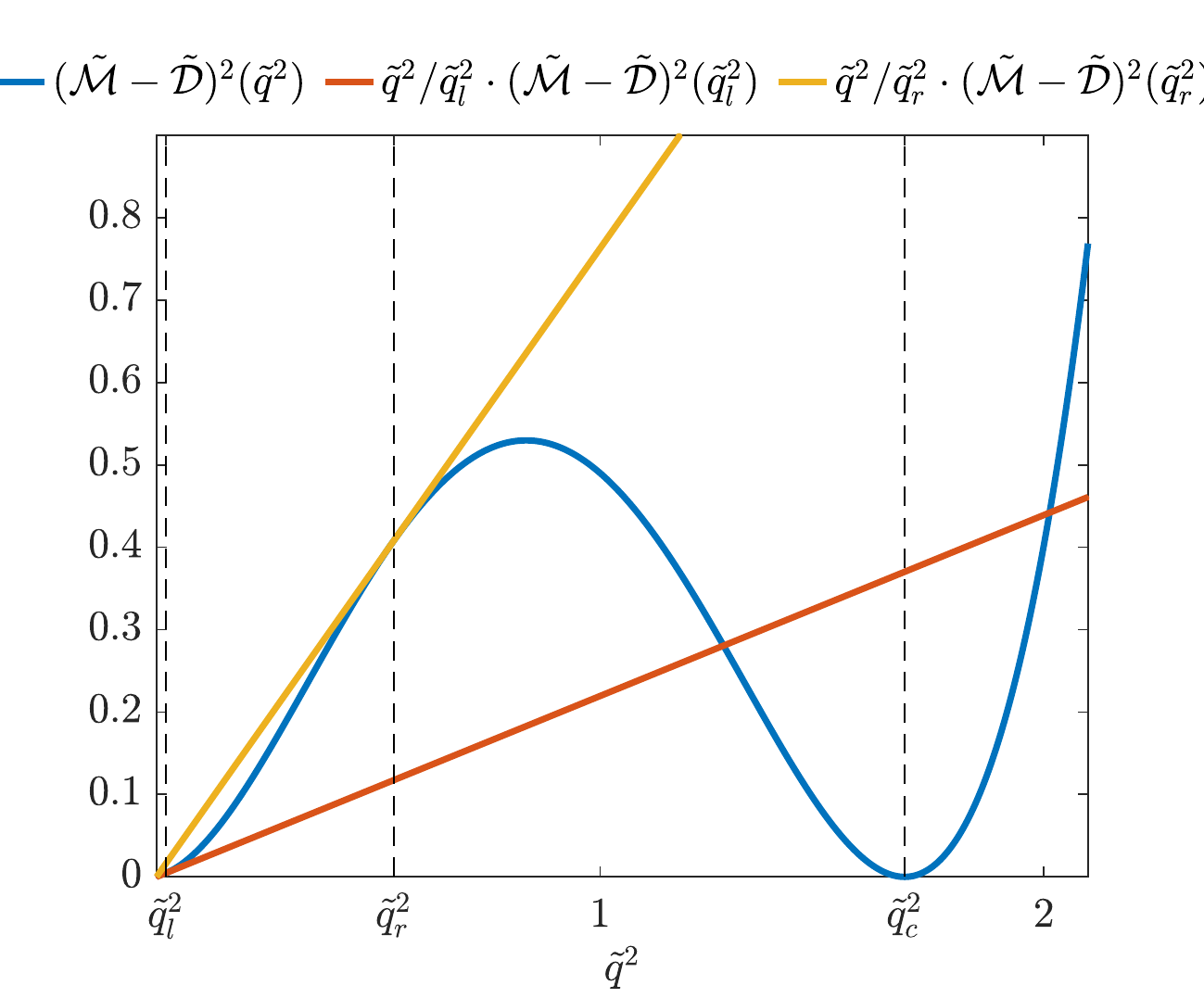}
  \caption{A plot of $(\tilde{\mathcal{M}}-\tilde{\mathcal{D}})^2(\tilde{q}^2)$ and two lines that pass through the origin and are tangent to the curve at $\tilde{q}_l^2$ and $\tilde{q}_r^2$. $\text{Da}=0.05$, $\alpha=1.5$.}
  \label{fig::Delta_demo}
\end{figure}

Fig.~\ref{fig::Delta_demo} shows a typical plot of $(\tilde{\mathcal{M}}-\tilde{\mathcal{D}})^2$ as a function of $\tilde{q}^2$ in blue while $\tilde{\mathcal{X}}$ as a function of $\tilde{q}^2$ is a line that passes through the origin whose slope is proportional to $\text{Pe}_\text{C}'$. The intersection of these two curves is where $\Delta=0$, and the region of $\Delta<0$ is where $(\tilde{\mathcal{M}}-\tilde{\mathcal{D}})^2$ is below the line $\tilde{\mathcal{X}}$. When $\text{Pe}_\text{C}'=0$, $\Delta \geq 0$. As $\text{Pe}_\text{C}'$ increases, the slope of $\tilde{\mathcal{X}}$ increases, the range of wavenumbers in which $\Delta<0$ expands.

Because $\Delta$ is a quartic polynomial of $\tilde{q}^2$, it has at most four roots. Now we would like to analyze the properties of its roots in order to determine the region of $\Delta<0$.

Notice that
$\tilde{\mathcal{M}}-\tilde{\mathcal{D}}$ has one valid root,
\begin{equation}
  \tilde{q}_c^2 = \frac{\alpha+1 +\sqrt{(\alpha+1)^2+4\alpha \text{Da}}}{2\alpha},
\end{equation}
as shown in Fig.~\ref{fig::Delta_demo}.
Note that $\tilde{q}_c^2 > (\alpha+1)/\alpha > 1$.
Hence, when $\text{Pe}_\text{C}'>0$, $(\tilde{\mathcal{M}}-\tilde{\mathcal{D}})^2$ and $\tilde{\mathcal{X}}$ has at least one intersection beyond $\tilde{q}_c^2$. This is shown graphically in Fig.~\ref{fig::Delta_demo}, where we see that any straight line that goes through the origin has one interaction with the blue curve at $\tilde{q}^2>\tilde{q}_c^2$. In other words, $\Delta$ has one root greater than $\tilde{q}_c^2$.

Since we are interested in the region $I_1>0$, or $\tilde{q}_- < \tilde{q} < \tilde{q}_+$, and we have $\tilde{q}_+ < 1 < \tilde{q}_c$, next, we focus on the roots of $\Delta$ within $[0,q_c]$. 
Because at $\tilde{q}=0$, $(\tilde{\mathcal{M}}-\tilde{\mathcal{D}})^2>0$ and $\tilde{\mathcal{X}}=0$, there is at least one root within $[0,q_c]$. Therefore, within this interval, there can be 1, 2, or 3 roots in total.

Having 2 roots in this interval or 3 roots in total for a quartic polynomial means that it has one root of multiplicity 2, or $\Delta=0$ and $d\Delta/d\tilde{q}^2=0$. Graphically, this corresponds to the line $\tilde{\mathcal{X}}$ being tangent to $(\tilde{\mathcal{M}}-\tilde{\mathcal{D}})^2$, as shown in Fig.~\ref{fig::Delta_demo}, where there are two solutions, and the root of multiplicity 2 is denoted as $\tilde{q}_l^2$ and $\tilde{q}_r^2$ respectively.
Mathematically it is equivalent to:
\begin{equation}
  \frac{(\tilde{\mathcal{M}}-\tilde{\mathcal{D}})^2}{\tilde{q}^2} = \frac{d\left((\tilde{\mathcal{M}}-\tilde{\mathcal{D}})^2\right)}{d\tilde{q}^2},
\end{equation}
or
\begin{equation}
  \frac{\tilde{\mathcal{M}}-\tilde{\mathcal{D}}}{\tilde{q}^2} = 2\frac{d(\tilde{\mathcal{M}}-\tilde{\mathcal{D}})}{d\tilde{q}^2}.
\end{equation}
Substituting in $\tilde{\mathcal{M}}$ and $\tilde{\mathcal{D}}$ [Eq.~\eqref{eqn::LSA_three_terms}], we obtain
\begin{equation}
  3\alpha q^4 - (1+\alpha) q^2 + \text{Da} = 0.
\end{equation}
A solution exists when
\begin{equation} \label{eqn::qlr_exists}
  \text{Da} \le \frac{(\alpha+1)^2}{12\alpha},
\end{equation}
and the roots are
\begin{equation}
  \tilde{q}^2_{l,r} = \frac{\alpha+1 \pm \sqrt{(\alpha+1)^2-12\alpha \text{Da}}}{6\alpha}.
\end{equation}
In the above equation, $\tilde{q}^2_l$ takes the minus sign and $\tilde{q}^2_r$ takes the plus sign.
In the discussion below, whenever we refer to $\tilde{q}_{l,r}$, we imply that the inequality in Eq.~\eqref{eqn::qlr_exists} holds.
The P\'{e}clet number $\text{Pe}_\text{C}'$ that corresponds to the tangent lines $\Delta(\tilde{q}^2_{l,r}) = 0$ is
\begin{eqnarray}
  \text{Pe}_{\text{C},l,r}' &=& \frac{(\tilde{\mathcal{M}}-\tilde{\mathcal{D}})^2(\tilde{q}_{l,r}^2)}{4 \alpha^{-1}\text{Da}\tilde{q}^2_{l,r}} = \frac{(\alpha\tilde{q}^4_{l,r}- (1+\alpha)\tilde{q}^2_{l,r} - \text{Da})^2}{4\alpha\text{Da} \tilde{q}^2_{l,r}} \nonumber \\
  &=& \frac{((\alpha+1)\tilde{q}_{l,r}^2 + 2\text{Da})^2}{9 \alpha \text{Da} \tilde{q}_{l,r}^2}.
\end{eqnarray}
When $0<\text{Pe}_\text{C}'<\text{Pe}_{\text{C},l}'$, $\Delta$ has one root $\tilde{q}_1$ in $(\tilde{q}_r,\tilde{q}_c)$ and $\Delta< 0$ in $(q_1,q_c]$.
When $\text{Pe}_{\text{C},l}<\text{Pe}_\text{C}'<\text{Pe}_{\text{C},r}$, $\Delta$ has three roots, $\tilde{q}_2 \in (0,\tilde{q}_l)$, $\tilde{q}_3 \in (\tilde{q}_l,\tilde{q}_r)$, and $\tilde{q}_4 \in (\tilde{q}_r,\tilde{q}_c)$, and $\Delta<0$ in $(q_2,q_3)$ and $(q_4,q_c]$.
When $\text{Pe}_\text{C}'>\text{Pe}_{\text{C},r}$, $\Delta$ has one root $\tilde{q}_5 \in [0,\tilde{q}_l]$, and $\Delta<0$ in $(q_5,q_c]$.
When $\text{Da}>(\alpha+1)^2/12\alpha$, $\tilde{q}_{l,r}$ does not exist and $\Delta$ has one root in $[0,\tilde{q}_c]$.

Recall that we are seeking the condition for unstable oscillatory modes, or $\Delta<0$ within the interval of $[\tilde{q}_-,\tilde{q}_+]$. Since the interval in which $\Delta<0$ expands with increasing $\text{Pe}_\text{C}'$, we need to find the critical condition that there exists $\tilde{q}^* \in [\tilde{q}_-,\tilde{q}_+]$ for which $\Delta(\tilde{q}^*)=0$ and for all $\tilde{q} \in [\tilde{q}_-,\tilde{q}_+]$, $\Delta(\tilde{q}) \geq 0$.
Therefore it is important to determine the order of $\tilde{q}_\pm$ and $\tilde{q}_{l,r}$. 

Setting $\tilde{q}_\pm = \tilde{q}_{l,r}$, we find that $\tilde{q}_-$ or $\tilde{q}_+$ is equal to $\tilde{q}_l$ or $\tilde{q}_r$ when
\begin{equation}
  \text{Da} = 1 -\frac{2}{\alpha}.
\end{equation}
Furthermore, we find that
when $\text{Da}<1-2/\alpha$, $\tilde{q}_l < \tilde{q}_- < \tilde{q}_r < \tilde{q}_+$. When $\text{Da}>1-2/\alpha$, the orders are: 
$\tilde{q}_l < \tilde{q}_- < \tilde{q}_+ < \tilde{q}_r$ when $1<\alpha<3$; $\tilde{q}_l < \tilde{q}_r < \tilde{q}_- < \tilde{q}_+$, when $3<\alpha<5$, and $\tilde{q}_- < \tilde{q}_l < \tilde{q}_r < \tilde{q}_+$ when $\alpha>5$.
Based on the analysis of the region of $\Delta<0$, we see that when $\tilde{q}_- < \tilde{q}_l < \tilde{q}_+$, the critical wavenumber $\tilde{q}^*$ can be $\tilde{q}_-$, $\tilde{q}_+$, or $\tilde{q}_l$, whichever makes $\Delta(\tilde{q}^*)=0$ at the smallest $\text{Pe}_\text{C}'$. In all other cases, including when $\tilde{q}_{l,r}$ do not exist ($\text{Da}>(\alpha+1)^2/12\alpha$), $\tilde{q}^*$ can only be $\tilde{q}_-$, $\tilde{q}_+$, whichever makes $\Delta(\tilde{q}^*)=0$ at the smallest $\text{Pe}_\text{C}'$.
Therefore, we define the P\'{e}clet number $\text{Pe}_\text{C}'$ that corresponds to $\Delta(\tilde{q}_\pm) = 0$,
\begin{multline}
  \text{Pe}_{\text{C},\pm}' = \frac{(\tilde{\mathcal{M}}-\tilde{\mathcal{D}})^2(\tilde{q}_\pm^2)}{4 \alpha^{-1}\text{Da}\tilde{q}^2_\pm}
  = \frac{(\tilde{q}_\pm^2 + \text{Da})^2}{  \alpha \text{Da} \tilde{q}_\pm^2}.
\end{multline}

In summary, when $\alpha>\alpha_\text{crit}$, the unstable modes become oscillatory when $\text{Pe}_\text{C}'>\text{Pe}_{\text{C},*}'$, where
\begin{equation}
  \label{eqn::critical_condition_oscillatory}
  \text{Pe}_{\text{C},*}' =
  \begin{cases}
    \text{min}\{\text{Pe}_{\text{C},+}', \text{Pe}_{\text{C},-}', \text{Pe}_{\text{C},l}'\} \; \text{when } \\ \qquad \text{Da}<(\alpha+1)^2/12\alpha, 
    \\ \qquad \text{Da}>1-2/\alpha, \\ \qquad \text{and } \alpha>5 \\
    \text{min}\{\text{Pe}_{\text{C},+}', \text{Pe}_{\text{C},-}'\} \, \text{otherwise}.
  \end{cases}
\end{equation}
Therefore, we have established that oscillatory instability occurs when chemoattractant diffusion is slow and chemotaxis is sufficiently fast.

Lastly, we note that by setting $\text{Pe}_{\text{C},+}' = \text{Pe}_{\text{C},-}'$, we find further that when $\alpha\text{Da} < 1$, $\text{Pe}_{\text{C},+}' < \text{Pe}_{\text{C},-}'$, and when $\alpha\text{Da} \geq 1$, $\text{Pe}_{\text{C},+}' \geq \text{Pe}_{\text{C},-}'$.

\section{Linear stability analysis in the $\text{Pe}_\text{R}-\phi_0$ phase diagram}
The results of linear stability analysis in Sec.~\ref{sec::SI_LSA} are described in terms of the three dimensionless parameters $\alpha$, $\text{Da}$, and $\text{Pe}_\text{C}'$. In their expressions (Eq.~\ref{eqn::dimensionless_parameters}), $\partial_\phi \tilde{\mu}_h$ is a function of $\phi_0$ and $\text{Pe}_\text{R}$, and $M$ is a function of $\phi_0$. $\phi_0$ and $\text{Pe}_\text{R}$ are the two dimensionless parameters in the conventional MIPS phase diagram. Therefore, in the main text, we define another version of the parameters that do not involve any dependence on $\phi_0$ and $\text{Pe}_\text{R}$: $\alpha_0$, $\text{Da}_0$ and $\text{Pe}_\text{C}$.
Using linear models for chemotactic sensing function and chemoattractant rate $f(\tilde{c}) = \tilde{c}$ and $g(\tilde{c}) = \tilde{c}$, the two versions of dimensionless parameters are related by
\begin{align} \label{eqn::dimensionless_parameters_conversion}
  \begin{split}
    \alpha &= - \phi_0 (\partial_\phi \tilde{\mu}_h) \alpha_0 \\
    \text{Da} &= -\frac{\phi_0}{\partial_\phi \tilde{\mu}_h} \text{Da}_0 \\
    \text{Pe}_\text{C}' &= \text{Pe}_\text{C} \frac{S}{k} \cdot \frac{(-1)}{\phi_0^2 \partial_\phi \tilde{\mu}_h}.
  \end{split}
\end{align}

Fig.~2 in the main text shows the chemotactic MIPS phase diagram in the plane of $\text{Pe}_\text{C} - \alpha_0$ at given $\text{Da}_0$, $\phi_0$ and $\text{Pe}_\text{R}$. The linear stability analysis results can be easily applied using the conversion in Eq.~(\ref{eqn::dimensionless_parameters_conversion}).

Fig.~1 in the main text shows the chemotactic MIPS phase diagram in the plane of $\text{Pe}_\text{R} -\phi_0$ at given $\alpha_0$, $\text{Da}_0$, and $\text{Pe}_\text{C}$. In this phase diagram, we would like to obtain the stability criteria and different types of instabilities expressed in terms of $\text{Pe}_\text{R}$ and $\phi_0$, which we derive in this section.

When there is no chemotaxis, the stability boundary is the spinodal curve $\partial_\phi \tilde{\mu}_h(\phi_0,\text{Pe}_{\text{R,sp}}) = 0$. Using Eq.~(\ref{eqn::dmu_h_dphi}), the spinodal curve can be written explicitly in terms of $\text{Pe}_\text{R}$:
\begin{equation} \label{eqn::Pe_R_sp}
  \text{Pe}_{\text{R,sp}} = \frac{\pi (\phi_0-\phi_m)^2 (\phi_0^{-1} - 2 - 0.6 \phi_0)  }{4 \phi_m (\phi_0 - 2\phi_m)}.
\end{equation}
Because $\partial_\phi \tilde{\mu}_h$ is linear with respect to $\text{Pe}_\text{R}$, in the following text, we give the stability and instability type conditions in terms of $\partial_\phi \tilde{\mu}_h$; the expression can then be easily written explicitly in terms of $\text{Pe}_\text{R}$.
Based on Eq.~(\ref{eqn::Pe_R_sp}), because $\text{Pe}_{\text{R,sp}}>0$ and $0<\phi_0<\phi_m$, we have $\phi_0^{-1} - 2 - 0.6 \phi_0<0$. Hence, we find that the spinodal curve spans the range of volume fractions given by $(-5+2\sqrt{10})/3 < \phi_0 < \phi_m$.

Based on Eq.~(\ref{eqn::critical_condition_Pe}), the criterion (1) ($\text{Pe}_\text{C}' \geq \text{Pe}_\text{C,crit}'$) can be written in terms $\text{Pe}_\text{C}$, $\text{Da}_0$, $\phi_0$, and $\partial_\phi \tilde{\mu}_h$ as
\begin{equation}
  \frac{S}{k} \cdot \text{Pe}_{\text{C}} \geq \phi_0^3
  \begin{cases}
    -\partial_\phi \tilde{\mu}_h / \phi_0, & \, \text{for } (-\mu'_h /\phi_0 \leq \text{Da}_0) \\
    \frac{(\text{Da}_0 - \partial_\phi \tilde{\mu}_h \phi_0^{-1})^2}{4\text{Da}_0}, & \text{for }\, (-\mu'_h /\phi_0 > \text{Da}_0)
  \end{cases},
\end{equation}
or explicitly in terms of $\partial_\phi \tilde{\mu}_h$:
\begin{equation}
  \label{eqn::boundary_1}
  -\frac{\partial_\phi \tilde{\mu}_h}{\phi_0} \leq
  \begin{cases}
    \frac{S \text{Pe}_\text{C}}{k \phi_0^3},  & \text{for }(\frac{S \text{Pe}_\text{C}}{k \phi_0^3} < \text{Da}_0) \\
    2\sqrt{\frac{S \text{Pe}_\text{C}}{k \phi_0^3} \text{Da}_0} - \text{Da}_0, & \text{otherwise}.
  \end{cases}
\end{equation}
Thus we have also obtained the expression for Boundary 1 in the $\text{Pe}_\text{R}-\phi$ phase diagram by setting Eq.~(\ref{eqn::boundary_1}) to equality, and the region above Boundary 1 satisfies criterion (1).

Similarly, criterion (2) ($\alpha \leq 1$ or $\text{Da} \geq (\alpha-1)^2/4\alpha$) can be written in terms of $\alpha_0$, $\text{Da}_0$, $\phi_0$, and $\partial_\phi \tilde{\mu}_h$ as
\begin{equation}
  \label{eqn::boundary_2}
  -\partial_\phi \tilde{\mu}_h  \leq \frac{1}{\phi_0 \alpha_0} \left( 1 + 2\phi_0 \sqrt{\text{Da}_0\alpha_0} \right).
\end{equation}
Thus we have also obtained the expression for Boundary 2 in the $\text{Pe}_\text{R}-\phi$ phase diagram by setting Eq.~(\ref{eqn::boundary_2}) to equality, and the region above Boundary 2 satisfies criterion (2).

Finite-wavelength instability $\text{Pe}_\text{C}'>1$ can be expressed as
\begin{equation}
  \label{eqn::I_II_boundary}
  - \partial_\phi \tilde{\mu}_h <  \frac{S \text{Pe}_\text{C}}{k \phi_0^2},
\end{equation}
which coincides with criterion (1) if $S \text{Pe}_\text{C}/(k \phi_0^3) < \text{Da}_0$. 
Setting Eq.~(\ref{eqn::I_II_boundary}) to equality gives the expression for the F/U boundary. Finite-wavelength instability exists between the F/U boundary and Boundary 1 when $S \text{Pe}_\text{C}/(k \phi_0^3) > \text{Da}_0$.


Finally, oscillatory instability occurs when $\text{Pe}_\text{C}'>\text{Pe}_{\text{C},*}'$ and $\alpha>\alpha_\text{crit}$.
The following derivation needs to be discussed separately depending on whether $\text{Da}<(\alpha+1)^2/12\alpha, \, \text{Da}>1-2/\alpha, \, \text{and } \alpha>5$, which is equivalent to
\begin{equation}
  \text{max}\{5,\phi_0 \sqrt{12\alpha_0\text{Da}_0} - 1\} < \alpha < 2 + \phi_0^2 \alpha_0 \text{Da}_0.
\end{equation}
If outside this region, the condition for oscillatory unstable mode is $\text{Pe}_\text{C}' > \text{min}\{\text{Pe}_{\text{C},+}', \text{Pe}_{\text{C},-}'\}$,
which is equivalent to
\begin{equation} \label{eqn::P_h}
  P \equiv 2\text{Pe}_\text{C} \frac{S}{k} \alpha_0^2 \text{Da}_0 \phi_0 > \min{\{h(u_+),h(u_-)\}},
\end{equation}
where
\begin{equation}
  h(u) = \frac{(u + 2\alpha\text{Da})^2}{u},
\end{equation}
and
\begin{equation}
  u_\pm = \alpha-1 \pm \sqrt{(\alpha-1)^2-4\alpha \text{Da}}.
\end{equation}
Because 
\begin{equation}
  h(u) \geq 8\alpha\text{Da},
\end{equation}
and the equality is attained at $u=2\alpha\text{Da}$,
this puts a lower bound on $\text{Pe}_\text{C}$:
\begin{equation}
  \label{eqn::Pec_lower_bound_oscillation}
  \text{Pe}_\text{C} \frac{S}{k} > \frac{4\phi_0}{\alpha_0}.
\end{equation}
Given the above constraint, the roots of $P=h(u)$ are
\begin{equation}
  u^\pm = \frac{1}{2} \left[ P - 4\alpha\text{Da} \pm \sqrt{P^2 - 8\alpha\text{Da}P } \right].
\end{equation}
Hence, Eq.~\eqref{eqn::P_h} is equivalent to $u^- < u_+ < u^+$ or $u^- < u_- < u^+$, that is, 
at least one of $u_\pm$ is in between the two roots $u^\pm$.
Recall that we would like to express the condition of oscillatory instability in terms of $\partial_\phi \tilde{\mu}_h$ explicitly. Because $\alpha\text{Da}=\alpha_0\text{Da}_0 \phi_0^2$, $u^\pm$ does not depend on $\partial_\phi \tilde{\mu}_h$.
But $u_\pm$ depends on $\partial_\phi \tilde{\mu}_h$ because of $\alpha$. Our goal, then, is to express $u^- < u_+ < u^+$ or $u^- < u_- < u^+$ explicitly in terms of $\alpha$, from which we obtain the condition in terms of $\partial_\phi \tilde{\mu}_h$ via $\partial_\phi \tilde{\mu}_h=-\alpha/(\alpha_0\phi_0)$.

To achieve the above goal, we first notice that
\begin{equation}
  \alpha - 1 = j(u_\pm) \equiv \frac{u_\pm^2 + 4\alpha \text{Da}}{2 u_\pm}.
\end{equation}
The following derivation uses the property that the minimum of $h(u)$ is obtained at $u=2\alpha\text{Da}$ and the minimum of $j(u)$ is obtained at $u=2\sqrt{\alpha\text{Da}}$.

We first consider condition (a) $2\sqrt{\alpha\text{Da}} \leq u^- < u^+$. Because $h(u^-)=P$, this requires: $2\alpha\text{Da} \geq 2\sqrt{\alpha\text{Da}}$, or $\alpha\text{Da} \geq 1$ (which corresponds to $\text{Pe}_{\text{C},+}'>\text{Pe}_{\text{C},-}'$ as noted in Sec.~\ref{sec::SI_oscillatory}), and $P \leq h(2\sqrt{\alpha\text{Da}})$, or
\begin{equation}
  \label{eqn::u-_or_u+}
  P \leq 2\sqrt{\alpha\text{Da}} (\sqrt{\alpha\text{Da}}+1)^2.
\end{equation}
Under condition (a), $u^- < u_+ < u^+$ or $u^- < u_- < u^+$ is equivalent to $j(u^-)<\alpha-1<j(u^+)$.

Next, we consider condition (b) $u^- < u^+ \leq  2\sqrt{\alpha\text{Da}}$, which requires $\alpha\text{Da}\leq 1$ and Eq.~\eqref{eqn::u-_or_u+}, then $u^- < u_+ < u^+$ or $u^- < u_- < u^+$ is equivalent to $j(u^+)<\alpha-1<j(u^-)$.

Lastly, if (c) $u^- <  2\sqrt{\alpha\text{Da}} < u^+$, which requires $P>h(2\sqrt{\alpha\text{Da}})$, or the opposite of Eq.~\eqref{eqn::u-_or_u+}, then $u^- < u_+ < u^+$ or $u^- < u_- < u^+$ is equivalent to $2\sqrt{\alpha\text{Da}}<\alpha-1<\max{\{j(u^-),j(u^+)\}}$, where the lower bound is the minimum of $j(u)$ ($\min_u{j(u)}=2\sqrt{\alpha\text{Da}}$).
Notice that $\alpha-1>2\sqrt{\alpha\text{Da}}$ is equivalent to $\text{Da}<(\alpha-1)^2/4\alpha$. Hence all conditions above imply $\text{Da}<\text{Da}_\text{crit}$ (or $\alpha>\alpha_\text{crit}$).

Now we have obtained the condition of oscillatory instability in terms of $\alpha$ and hence $\partial_\phi \tilde{\mu}_h$ explicitly, which has an upper and lower bound. The upper bound coincides with or is below Boundary 2.




In summary, in this section, we have derived the conditions for stability or instability (both type F/U and type S/O) in the $\text{Pe}_\text{R}-\phi_0$ phase diagram, by expressing them explicitly in terms of $\partial_\phi \tilde{\mu}_h$.

\section{Numerical simulations}
In this section, we describe the details of numerical simulations. Firstly, we define the characteristic length scale to be $l_0 \equiv \sqrt{\kappa} \sim U_0 \tau_R$, which is on the order of the persistence length \cite{Stenhammar2013,Cates2015}. We define the characteristic time scale to be $t_0 \equiv \kappa/M_0 \sim \tau_R$, which is on the order of the ABP reorientation time.
The characteristic length and time scales motivate us to define the characteristic velocity $u_0 \equiv l_0 / t_0 \sim U_0$, which we will use in Sec. \ref{sec::SI_oscillatory}.

All simulations in this work are performed in a periodic domain of size $[100l_0,100l_0]$.
The governing equations are solved using the finite volume method to ensure conservation of particle volume fraction and chemoattractant concentration. We use an implicit solver of variable order as the time-stepper with adaptive time stepping, adaptive order and error control~\cite{Shampine1997}. Simulations are solved on a grid of size $[256,256]$.
The initial condition for $\phi(x,t=0)$ is the homogeneous state $\phi_0$ with added spatially uncorrelated Gaussian noise at each grid point with a standard deviation of 0.02. The initial condition for the chemoattractant concentration is the homogeneous state $\tilde{c}(x,t=0)=\tilde{c}_0=S/(k\phi_0)$ which satisfies the steady state condition.

Snapshots in Fig.~1 and Fig.~2 in the main text are taken at $t=4\times10^4 t_0$ and $t=2\times10^4 t_0$, respectively.
Note that in Fig.~2(b-c), of the two eigenvalues, only the higher one $\tilde{\omega}_+$ is shown since it determines the stability.
In all phase diagrams where simulations are displayed (Fig.~1-2, SI Movie~1-7), the parameters for the simulations correspond to the coordinates of the center of the images.

\section{Characterization of coarsening dynamics}
\label{sec::SI_coarsen}
\begin{figure}[h]
  \includegraphics[width=\columnwidth]{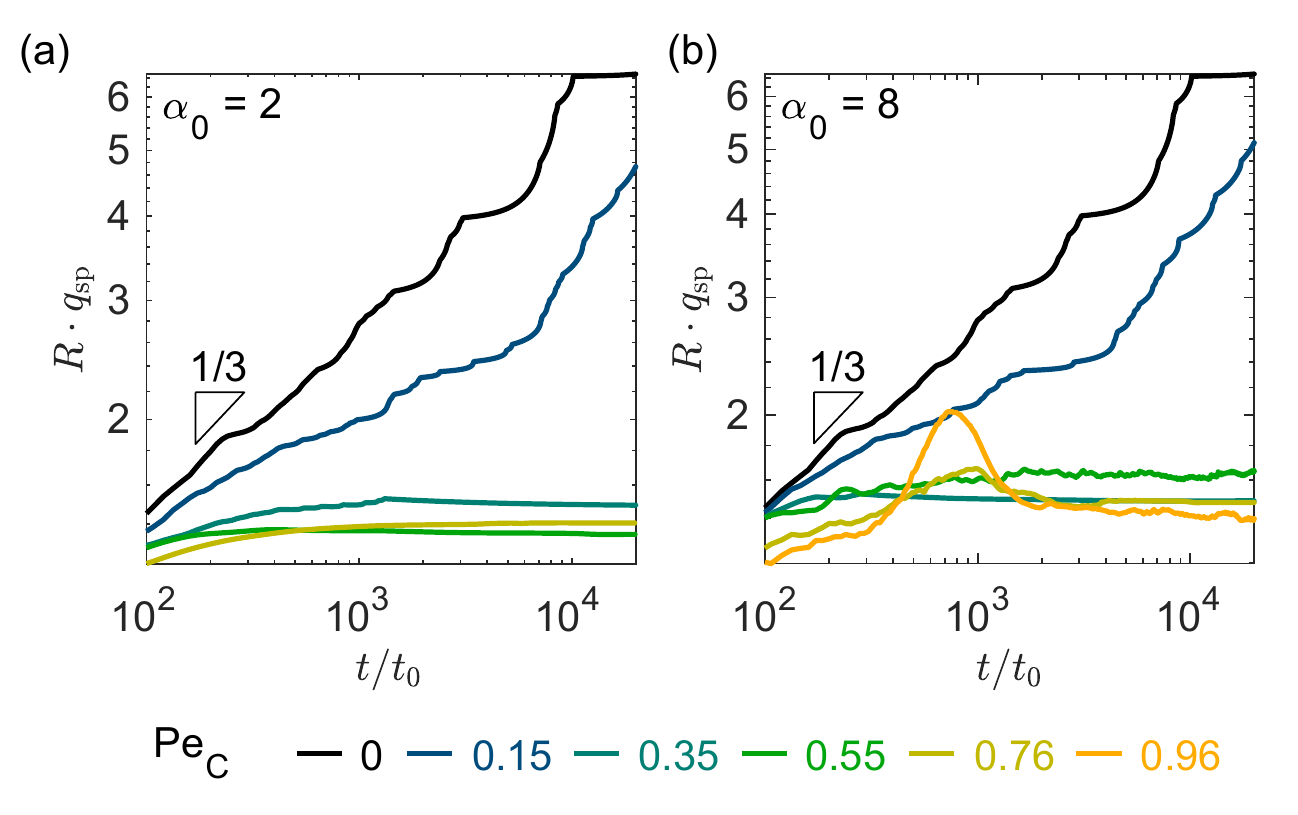}
  \caption{Evolution of the characteristic domain sizes that correspond to the simulations at $\alpha_0=2$ and $8$ and increasing values of $\text{Pe}_\text{C}$ in Fig.~2(a) in the main text.}
  \label{fig::coarsen_from_main}
\end{figure}

It is known that in conventional MIPS the size of phase-separated domain coarsens over time~\cite{Stenhammar2013}. However, as shown in the main text, chemotaxis can arrest such coarsening. In this section, we quantify the coarsening dynamics by plotting the evolution of the characteristic domain size over time, defined to be~\cite{Furukawa2000,Laradji1996,Mao2019}
\begin{equation}
  R(t)=\left[ \frac{\int{|\mathbf{q}|\mathcal{S}(\mathbf{q},t)d\mathbf{q}}}{\int{\mathcal{S}(\mathbf{q},t)d\mathbf{q}}} \right]^{-1},
\end{equation}
where $\mathcal{S}(\mathbf{q},t)$ is the structure factor associated with spatial variations in particle volume fraction
\begin{equation}
  \mathcal{S}(\mathbf{q},t) = |\Delta \hat{\phi}(\mathbf{q},t)|^2,
\end{equation}
where $\Delta \hat{\phi}$ is the Fourier transform of $\Delta\phi = \phi - \phi_0$.

Fig.~\ref{fig::coarsen_from_main}(a-b) shows the normalized characteristic domain size $Rq_\text{sp}$ with respect to time that correspond to the simulations with $\alpha_0=2$ and $8$ in Fig.~2(a) in the main text ($\text{Da}_0=0.5$, $\text{Pe}_\text{R}=10^{-3}$, $\phi = 0.8$).
The domain size of the case of non-chemotactic MIPS ($\text{Pe}_\text{C}=0$) grows as $R\sim t^{1/3}$ (black curve), consistent with the growth law of spinodal decomposition~\cite{Bray2002}, showing that the coarsening persists.
With increasing $\text{Pe}_\text{C}$, this coarsening slows down (blue to chartreuse curves), eventually becomes arrested and gives rise instead to finite-sized domains characteristic of a Type F instability.

\begin{figure}[h]
  \includegraphics[width=\columnwidth]{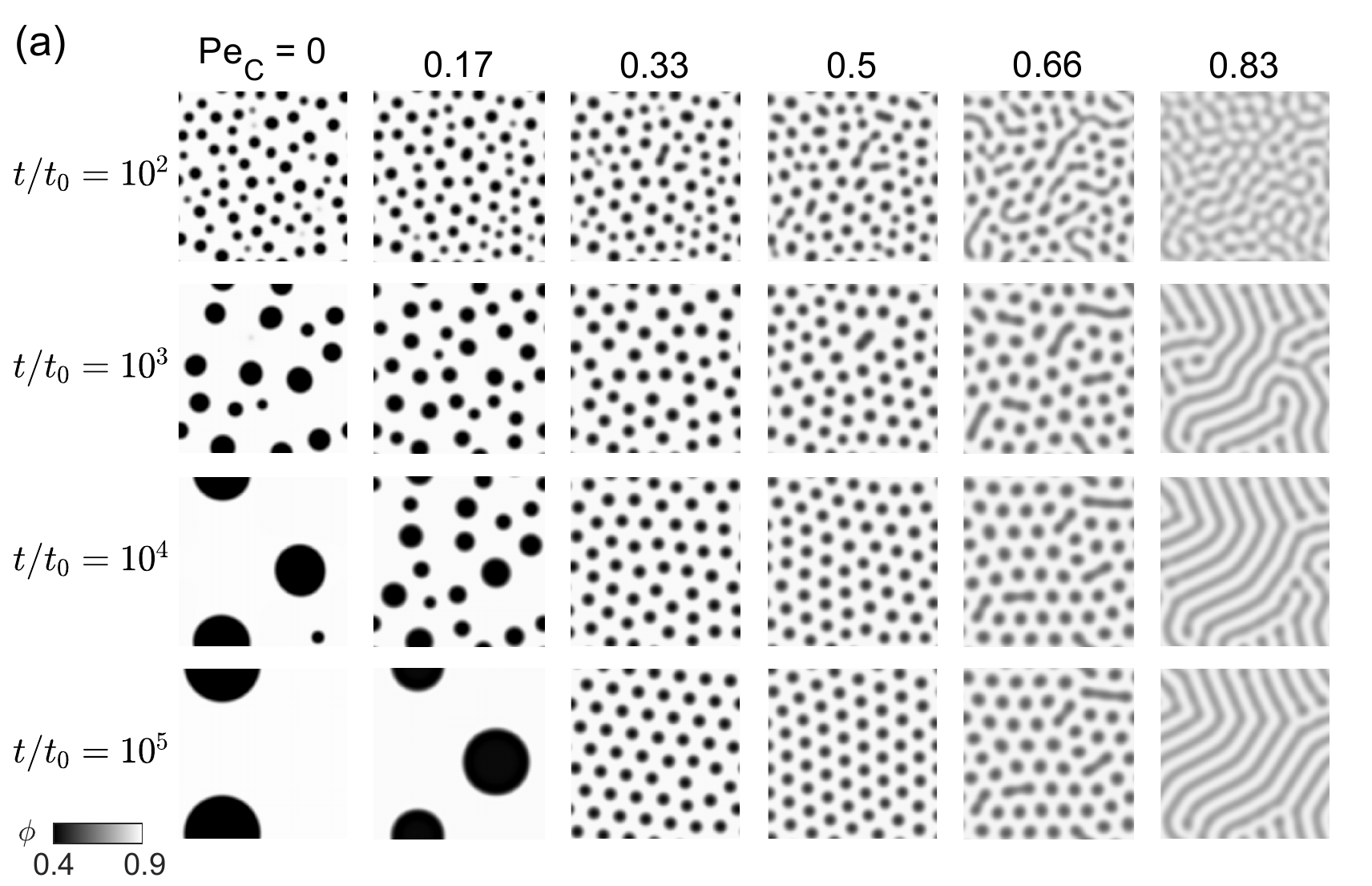}
  \includegraphics[width=\columnwidth]{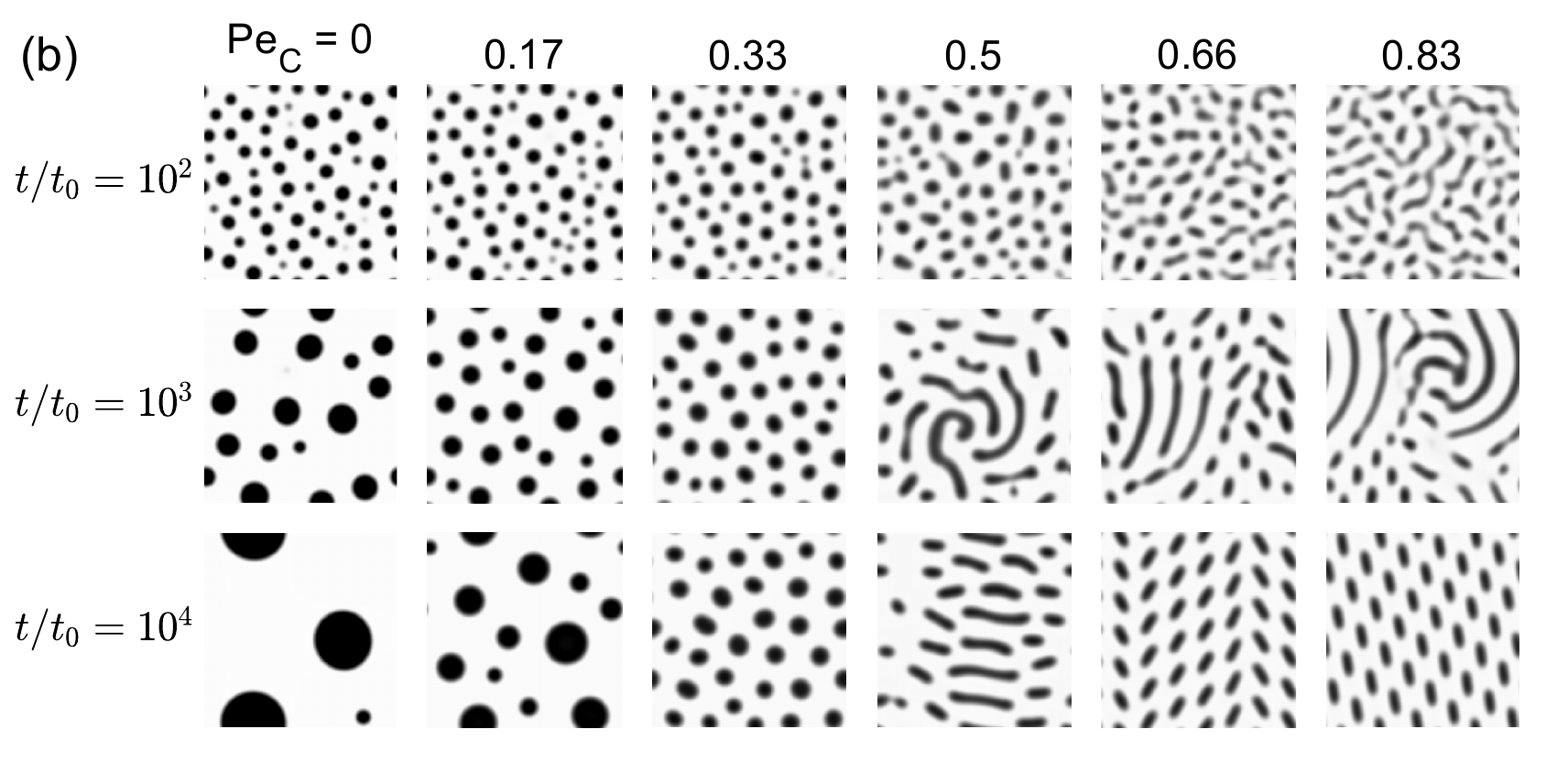}
  \includegraphics[width=\columnwidth]{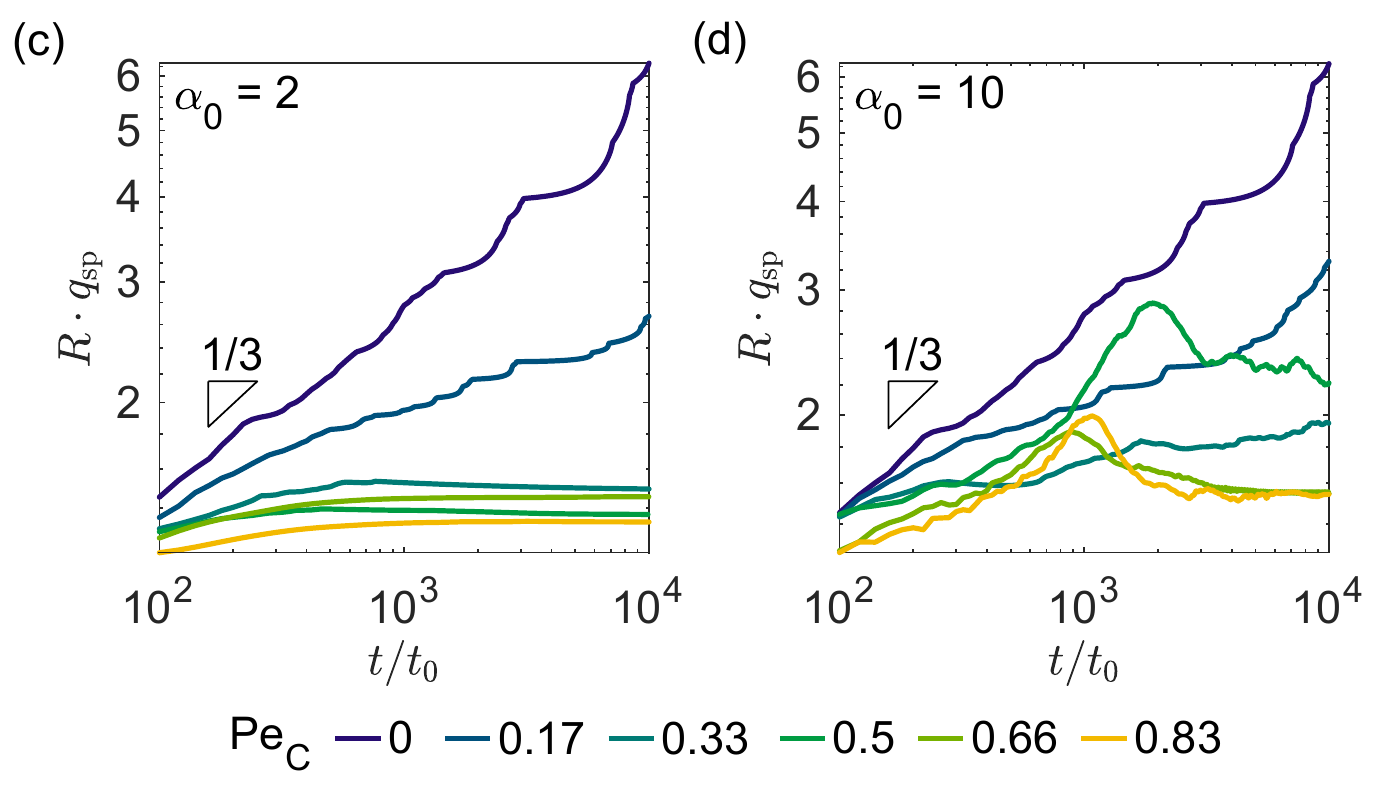}
  \caption{(a-b)~Snapshots of simulations and (c-d)~evolution of the characteristic domain size $R(t)$  at $\text{Da}_0=0.5$, $\text{Pe}_\text{R}=10^{-3}$, $\phi_0=0.8$, $\alpha_0=2$~(a,c) and 10~(b,d), and increasing values of $\text{Pe}_\text{C}$.}
  \label{fig::coarsen_1}
\end{figure}

Next, we show more examples of the coarsening dynamics with smaller steps of increasing $\text{Pe}_\text{C}$.
Fig.~\ref{fig::coarsen_1} shows the snapshots of the coarsening process and the dependence on $\text{Pe}_\text{C}$ when $\text{Da}_0=0.5$, $\text{Pe}_\text{R}=10^{-3}$, $\phi = 0.8$ and $\alpha_0=2$ and 10, along with the corresponding $R(t)$. $\text{Pe}_\text{C}$ is chosen such that it is equally spaced between 0 and 90\% of the critical $\text{Pe}_\text{C}$ that corresponds to Boundary 1.
We see that when the patterns are stationary ($\alpha_0=2$), with increasing $\text{Pe}_\text{C}$, the coarsening generally slows down, and the domain size at steady state decreases. For $\alpha_0=10$, coarsening also slows down with increasing $\text{Pe}_\text{C}$ until the pattern becomes oscillatory, when initially, the coarsening may be faster than stationary patterns. Note that $R(t)$ is typically non-monotonic for oscillatory patterns, and when $R(t)$ converges to a steady value at longer time, the steady value decreases with increasing $\text{Pe}_\text{C}$.

\begin{figure}[h]
  \includegraphics[width=\columnwidth]{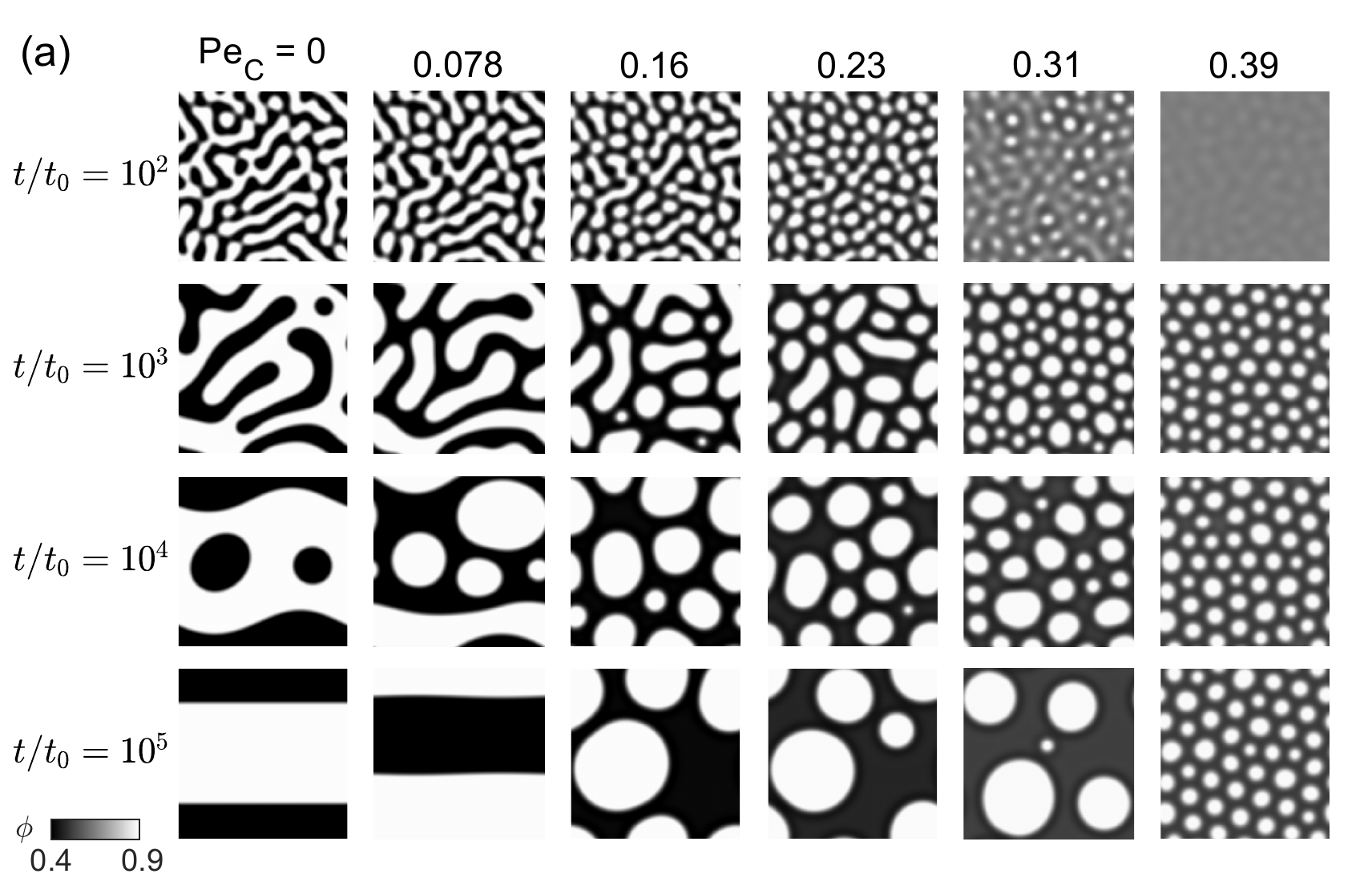}
  \includegraphics[width=\columnwidth]{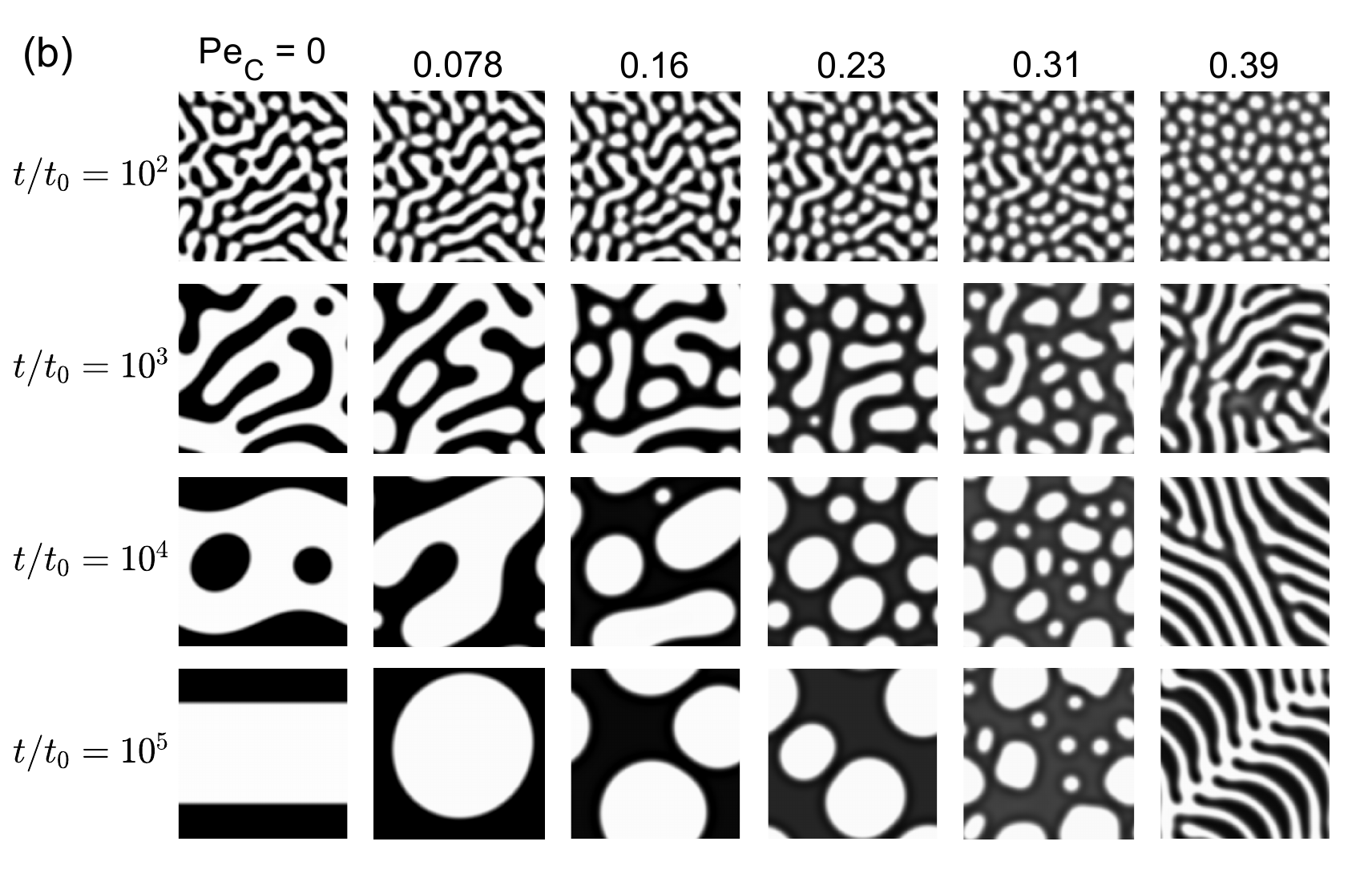}
  \includegraphics[width=\columnwidth]{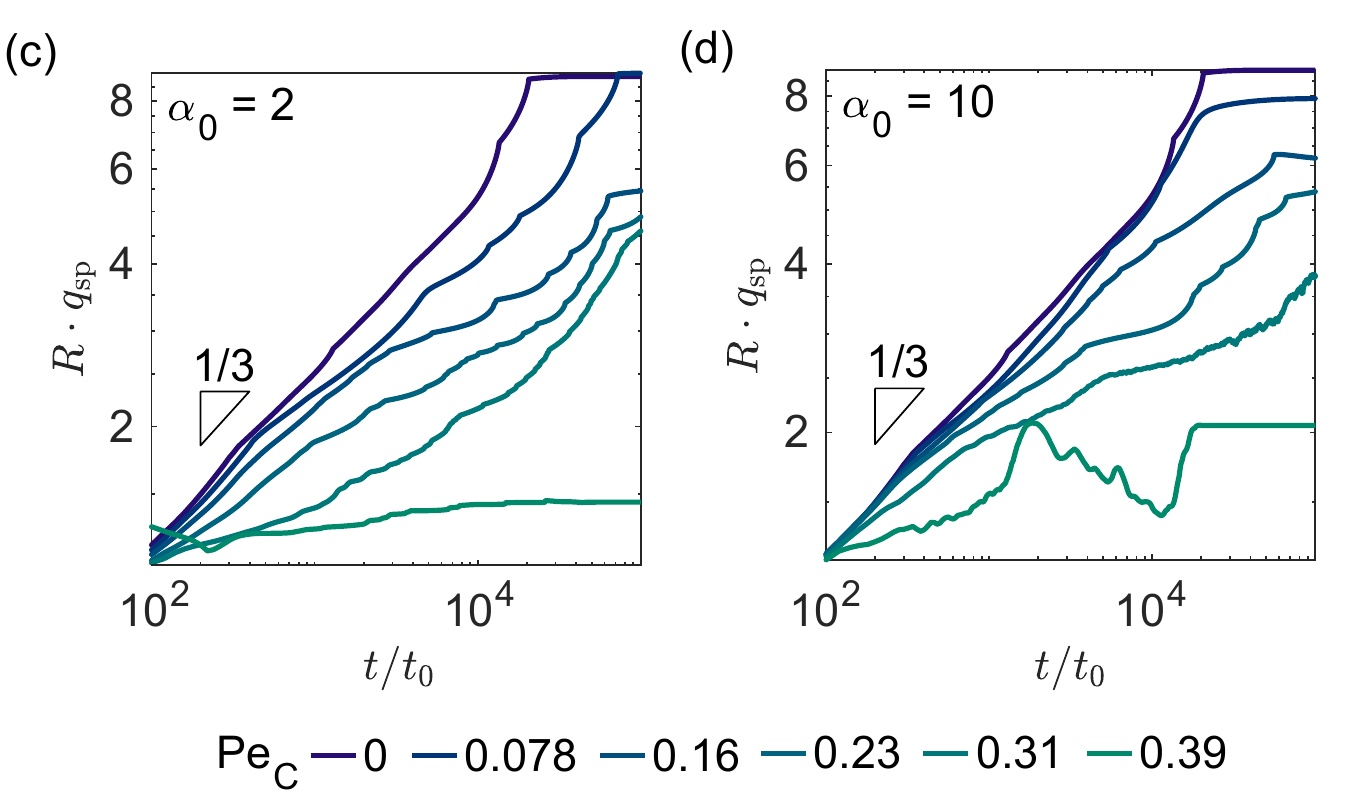}
  \caption{(a-b)~Snapshots of simulations and (c-d)~evolution of the characteristic domain size $R(t)$ at $\text{Da}_0=0.5$, $\text{Pe}_\text{R}=10^{-3}$, $\phi_0=0.65$ for $\alpha_0=2$~(a,c) and 10~(b,d).}
  \label{fig::coarsen_2}
\end{figure}

The observation above can also be seen in Fig.~\ref{fig::coarsen_2}, which shows the results for $\phi_0 = 0.65$ (other dimensionless parameters are identical to Fig.~\ref{fig::coarsen_1}).
Because the pattern for $\phi_0=0.65$ is bicontinuous for a larger fraction of the time, at $\text{Pe}_\text{C}=0$, the growth curve $R(t)$ is smoother than $\phi_0=0.8$ shown in Fig.~\ref{fig::coarsen_from_main}, which shows step increase due to events of dissolution and merger of phases.
Again, we confirm the $R\sim t^{1/3}$ power law for non-chemotactic MIPS ($\text{Pe}_\text{C}=0$).
In the main text, we referred the readers to Figs.~\ref{fig::coarsen_1} and \ref{fig::coarsen_2} for snapshots of the non-chemotactic spinodal decomposition.

Fig.~\ref{fig::coarsen_2} also shows that the slope of $\ln{R}-\ln{t}$ decreases with increasing $\text{Pe}_\text{C}$ for stationary patterns at both $\alpha_0=2$ and $10$, again indicating slower coarsening.
A traveling pattern is instead observed at $\alpha_0=10$ and $\text{Pe}_\text{C}=0.39$, for which $R(t)$ shows non-monotonic behavior.

\section{Characeterization of small-amplitude fluctuation}
\label{sec::SI_dispersion}
\begin{figure}
  \includegraphics[width=\columnwidth]{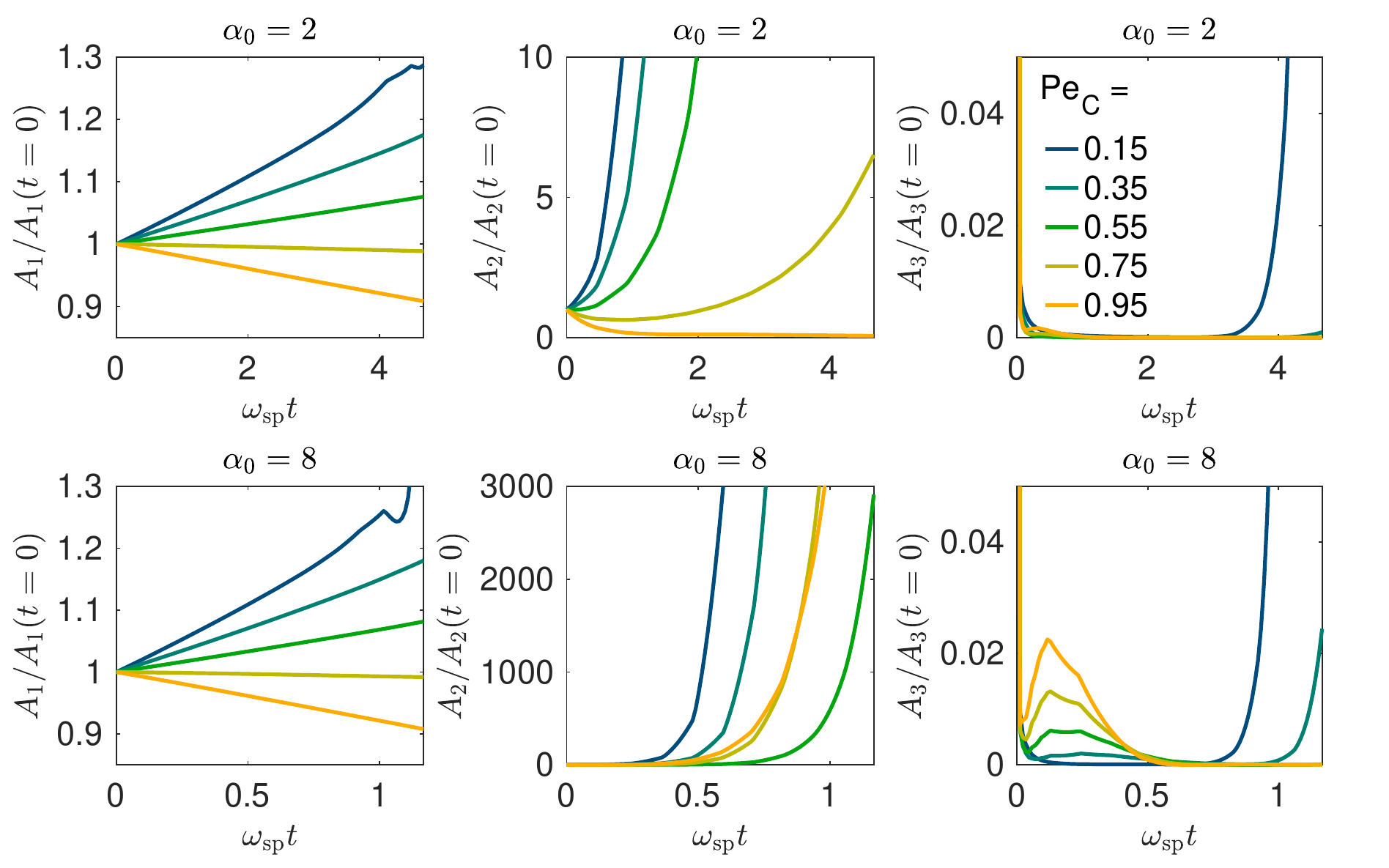}
  \caption{The time evolution of the perturbation in the long ($A_1$), medium ($A_2$) and short wavelength ($A_3$) regimes. The parameters correspond to the columns of $\alpha_0=2$ and $\alpha_0=8$ in Fig. 2(a) in the main text.}
  \label{fig::dispersion_sim}
\end{figure}

In this section, we verify the classification of type F/U instability based on the linear stability analysis using numerical simulations.
To compare with the dispersion relation shown in Fig.~2(b-c) in the main text, we perform simulations at these parameters. The initial condition for $\phi(x)$ is a homogeneous $\phi_0$ with added spatially uncorrelated Gaussian noise at each grid point with a standard deviation of 0.001. We use a small amplitude perturbation here to reduce the nonlinear effect. The initial condition for the chemoattractant concentration is the homogeneous state $\tilde{c}(x,t=0)=\tilde{c}_0=S/(k\phi_0)$ with added noise that has the opposite sign as the added noise for $\phi(x)$.
We observe the early time evolution of long, medium, and short wavelength modes by defining the following quantities based on the structure factor: $A_1(t) = \int_0^{0.16q_\text{sp}}{S(\mathbf{q},t)d\mathbf{q}}$, $A_2(t) = \int_{0.5q_\text{sp}}^{q_\text{sp}}{S(\mathbf{q},t)d\mathbf{q}}$, $A_3(t) = \int_{\sqrt{2}q_\text{sp}}^{\infty}{S(\mathbf{q},t)d\mathbf{q}}$.
Fig.~\ref{fig::dispersion_sim} shows that in the long wavelength regime for both $\alpha_0=2$ and $\alpha_0=8$, the amplitude of the perturbation decreases for $\text{Pe}_\text{C}=0.75$ and $0.95$ and increases for other cases, consistent with the dispersion relation in Fig.~2(b-c), verifying that $\text{Pe}_\text{C}=0.75$ and $\text{Pe}_\text{C}=0.95$ correspond to type F and other cases correspond to type U.
In the medium wavelength regime where the instability grows the fastest, the perturbation grows for all cases except for $\alpha_0=2$ and $\text{Pe}_\text{C}=0.95$, which is linearly stable at all wavelengths. Note that the curve is nonmotonic for $\alpha_0=2$ and $\text{Pe}_\text{C}=0.75$. This can be due to the coupling between $\phi$ and $c$, since the perturbation we impose is not an eigenvector in the linear stability analysis.
In the short wavelength regime, initially all amplitudes decrease sharply.

\section{Characterization of oscillatory pattern formation}
\label{sec::SI_oscillatory}
In Fig.~2(a) in the main text, we plot the velocity (red arrows) of the patterns. In this section, we show the definition of the velocity and its dependence on $\text{Pe}_\text{C}$ and $\alpha_0$.

To quantify the velocity of stripes that span the entire domain and spirals, we define a level set velocity $\mathbf{u}$, that is, the velocity at which contours of $\phi$ move in the direction along the gradient: $\mathbf{u}=-\partial_t \phi \cdot \nabla \phi / |\nabla\phi|^2$. Note that the level set velocity is undefined when the gradient vanishes.
Cases that use this definition are: $\text{Pe}_\text{C}=0.96$, $\alpha_0=4,6,8,10$, and $\text{Pe}_\text{C}=0.76$, $\alpha_0=10$. For all other cases, which exhibit dot-like and short stripe-like patterns, $\mathbf{u}$ is instead defined to be the velocity of the center of mass of each of the disjoint regions defined by $\{x|\phi(x)<0.7\}$ to facilitate ease of visualization. The vectors of $\mathbf{u}$ are indicated by the red arrows in Fig.~2(a). The scale bar $u_0 \equiv l_0/t_0 \sim U_0$ indicates the characteristic velocity.

\begin{figure}[h]
  \includegraphics[width=0.6\columnwidth]{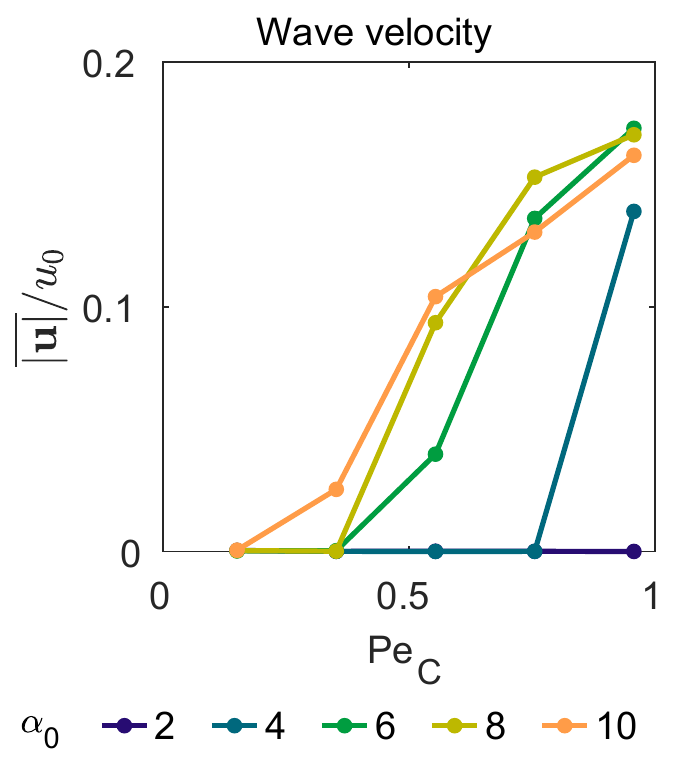}
  \caption{Average speed of patterns corresponding to Fig.~2(a) in the main text.}
  \label{fig::wave_velocity}
\end{figure}

Fig.~\ref{fig::wave_velocity} summarizes the velocity of patterns shown in Fig.~2(a). We compute the average speed $\overline{\mathbf{u}}$ at $t/t_0=2\times 10^4$. The level set velocity is averaged over all grid points for which $|\nabla\phi| >0.03l_0^{-1}$ to avoid inaccuracy when the magnitude of the gradient is small (level set velocity is undefined when $|\nabla\phi|=0$), while the center of mass velocity is averaged over all disjoint regions. We see that patterns move faster with increasing $\text{Pe}_\text{C}$ and the onset of motion occurs at lower $\text{Pe}_\text{C}$ with increasing $\alpha_0$.

As a reference, the average speed of the patterns at $\text{Pe}_\text{C}=0.95$ is on the order of $0.1u_0$, which means that the pattern travels at a speed that is an order of magnitude smaller than the speed at which ABP particles self-propel.

\section{Application in living and synthetic systems}
\label{sec::SI_application}
In this section, we estimate the values of the dimensionless parameters of living systems and discuss ways to study the phase diagram of chemotactic MIPS experimentally by tuning certain properties of synthetic colloidal systems.

\textbf{\emph{Populations of motile bacteria.}} We use \textit{Myxococcus xanthus} and \textit{Escherichia coli} as representative examples to draw estimates of parameter values from. $\text{Pe}_\text{R}\sim10^{-2}$, and hence, cells may undergo MIPS at sufficiently high cell density~\cite{Liu2019}. We therefore take $\partial_\phi \tilde{\mu}_h \sim 1$ as shown in Eq.~\eqref{eqn::dmu_upper_bound} for cells in the spinodal region of MIPS. The experimentally measured diffusivity $M_0$ can range from $\sim0.1$ to $\sim10^2~\upmu\text{m}^2/s$~\cite{Alert2022,Liu2019,Fu2018}, typically lower than chemoattractant diffusivity, which suggests that $\alpha \lesssim 1$. The typical chemoattractant depletion length is $\sqrt{D_c/k}\sim1~\upmu\text{m}$~\cite{Alert2022}, and persistence length $l_0$~\cite{Cates2012} is about $20~\upmu\text{m}$, hence $\text{Da}\sim 10^2$. These estimates suggest that populations of motile bacteria satisfy criterion (2), indicating that MIPS can be suppressed by chemotaxis when $\text{Pe}_\text{C}' \approx \text{Pe}_\text{C} \cdot S/k $ is sufficiently large. Because $\text{Pe}_\text{C}\sim10$~\cite{Alert2022,Fu2018,Bhattacharjee2021}, when chemoattractant is abundant ($S$ is large), MIPS is suppressed. Conversely, when chemoattractant is limited, we expect that MIPS can occur.

For synthetic systems such as self-propelled colloids, because $\text{Da} \sim U_0^2 \tau_R^2 k / D_c$, and $\alpha \sim U_0^2 \tau_R / D_c$,
$\text{Da}$ and $\alpha$ can be tuned via the swimming velocity $U_0$ e.g., using external stimuli such as light~\cite{Buttinoni2012,Stenhammar2016,Arlt2018,Frangipane2018}. In addition, $\text{Da}$ can be tuned by changing the reactive material to alter the chemoattractant uptake rate $k$.
Finite-sized domains arise experimentally if synthetic chemotactic colloids have a low uptake rate $k$, which leads to smaller $\text{Da}$. With a smaller $\text{Da}$ and a larger $\alpha$ (such as by increasing $U_0$), oscillatory dynamics involving clusters of colloidal particles traveling in space may arise.


\begin{table*}
    \centering
    \begin{tabular}{c|l}
       \textbf{Variable} & \textbf{Physical meaning} \\ \hline
       $a$ & ABP particle radius \\ \hline
       $U_0$ & ABP self-propulsion speed \\ \hline
       $\tau_R$ & ABP reorientation time \\ \hline
       $M_0 = \frac{1}{2}U_0^2 \tau_R$ &  ABP active diffusivity \\ \hline
       $\phi_0$ & Average ABP volume fraction \\ \hline
       $\tilde{\mu}_h$ & Normalized ABP chemical potential \\ \hline
       $l_0 \equiv \sqrt{\kappa} \sim U_0 \tau_R$ & Characteristic length scale of the width of the MIPS interface \\ \hline
       $D_c$ & Chemoattractant diffusivity \\ \hline
       $k$ & Chemoattractant uptake rate coefficient \\ \hline
       $S$ & Chemoattractant supply rate \\ \hline
       $\chi_0$ & Chemotactic coefficient \\ \hline
       $f(\tilde{c})=\tilde{c}$ & Chemotactic sensing function \\ \hline
       $g(\tilde{c})=\tilde{c}$ & Dependence of chemoattractant uptake rate on chemoattractant concentration \\ \hline
       $\text{Pe}_\text{R} \equiv \frac{a}{U_0\tau_R}$ & Reorientational P\'{e}clet number: directedness of ABPs \\ \hline
       $\alpha_0 \equiv \frac{M_0}{D_c}$ & Ratio of single-particle ABP to chemoattractant diffusivity \\ \hline 
       $\text{Da}_0 \equiv \frac{\kappa k}{D_c}$ & Chemoattractant uptake rate to diffusion rate over $\sqrt{\kappa}$ \\ \hline
       $\text{Pe}_\text{C} \equiv \frac{\chi_0}{M_0}$ & Chemotactic P\'{e}clet number: ratio of ABP chemotactic coefficient to diffusivity \\ \hline
       $\alpha \equiv \frac{-M_0\phi_0 \partial_\phi \tilde{\mu}_h}{D_c} = -\alpha_0 \phi_0 \partial_\phi \tilde{\mu}_h $ & Ratio of effective collective ABP to chemoattractant diffusivity \\ \hline
       $\text{Da} \equiv -\frac{\kappa k\phi_0 g'}{D_c \partial_\phi \tilde{\mu}_h} = -\text{Da}_0 \frac{\phi_0 g'}{\partial_\phi \tilde{\mu}_h}$ & Damk{\"o}hler number: effective chemoattractant uptake to diffusion rate \\ \hline
       $\text{Pe}_\text{C}' \equiv -\frac{\chi_0}{M_0\phi_0\partial_\phi \tilde{\mu}_h} \frac{f' g}{g'} = -\text{Pe}_\text{C} \frac{f'g}{\phi_0\partial_\phi \tilde{\mu}_h g'}$ & Reduced chemotactic P\'{e}clet number: effective ABP chemotactic to diffusivity rate
       
    \end{tabular}
    \caption{Summary of variables and dimensionless parameters.}
    \label{table::variables}
\end{table*}

\section{Supplementary movies \\    (available upon request)}
\begin{enumerate}
    \item Animated profiles of $\phi(\mathbf{x})$ that show non-chemotactic MIPS in the $\text{Pe}_\text{R}-\phi_0$ phase diagram ($\text{Pe}_\text{C}=0$).
    \item Animated profiles of $\phi(\mathbf{x})$ for the simulations in Fig.~1(b). $\text{Da}_0=0.2$, $\alpha_0=1$, $\text{Pe}_\text{C}=1$.
    \item Animated profiles of $\phi(\mathbf{x})$ for the simulations in Fig.~1(d). $\text{Da}_0=0.2$, $\alpha_0=4$, $\text{Pe}_\text{C}=1$.
    \item Animated profiles of $\phi(\mathbf{x})$ for the simulations in Fig.~1(f). $\text{Da}_0=0.5$, $\alpha_0=10$, $\text{Pe}_\text{C}=0.35$.
    \item Animated profiles of $c(\mathbf{x})$ for the simulations in Fig.~1(b). $\text{Da}_0=0.2$, $\alpha_0=1$, $\text{Pe}_\text{C}=1$.
    \item Animated profiles of $c(\mathbf{x})$ for the simulations in Fig.~1(d). $\text{Da}_0=0.2$, $\alpha_0=4$, $\text{Pe}_\text{C}=1$.
    \item Animated profiles of $\phi(\mathbf{x})$ for the simulations in Fig.~2(a). $\text{Da}_0=0.5$, $\text{Pe}_\text{R}=10^{-3}$, $\phi_0=0.8$.
    
\end{enumerate}

\end{document}